\documentclass{aa}  

\usepackage{graphicx}
\usepackage{txfonts}
\usepackage{array}
\usepackage{caption}
\usepackage{booktabs} 

\newcommand{\ch}[1]{{$\rm #1$}}
\newcommand{\hl}[1]{{{#1}}}
\newcommand{\hll}[1]{{{#1}}}
\newcommand{\pdm}{{P{\small RO}D{\small I}M{\small O}}\xspace}
\usepackage{lipsum}
\usepackage[hidelinks,colorlinks=true,linkcolor=blue,anchorcolor=black,citecolor=blue,filecolor=black,menucolor=black,runcolor=black,urlcolor=black]{hyperref}
\usepackage{placeins}

\begin{document}

   \title{Molecular diagnostics for the mid-infrared emission of planet-forming disks}

   \subtitle{Carbon and oxygen elemental abundances}

   \author{Aditya M. Arabhavi
          \inst{1}
          \and
          Inga Kamp \inst{1}
          \and
          Ewine F. van Dishoeck \inst{2,3}
          \and
          Peter Woitke \inst{4}
          \and
          Christian Rab \inst{5,6}
          \and
          Wing-Fai Thi \inst{3}
          \and
          Till Kaeufer \inst{7}
          \and
          Jayatee Kanwar \inst{8}
          \and
          Beno\^{i}t Tabone \inst{9}
          \and
          Pac\^{o}me Esteve \inst{9}
          \and
          Marissa Vlasblom \inst{2}
          }

   \institute{
            Kapteyn Astronomical Institute, Rijksuniversiteit Groningen, Postbus 800, 9700AV Groningen, The Netherlands\\
              \email{arabhavi@astro.rug.nl}
        \and
             Leiden Observatory, Leiden University, 2300 RA Leiden, the Netherlands
        \and 
            Max-Planck Institut f\"{u}r Extraterrestrische Physik (MPE), Giessenbachstr. 1, 85748, Garching, Germany
        \and
            Space Research Institute, Austrian Academy of Sciences, Schmiedlstrasse 6, A-8042 Graz, Austria
        \and
            University Observatory, Faculty of Physics, Ludwig-Maximilians-Universität München, Scheinerstr. 1, D-81679 Munich, Germany
        \and
            Max-Planck-Institut für extraterrestrische Physik, Giessenbachstrasse 1, D-85748 Garching, Germany
        \and
            Department of Physics and Astronomy, University of Exeter, Exeter EX4 4QL, UK
        \and
            Department of Astronomy, University of Michigan, 1085 S. University, Ann Arbor, MI 48109, USA
        \and
            Universit\'e Paris-Saclay, CNRS, Institut d’Astrophysique Spatiale, 91405, Orsay, France
             }

   \date{Received ***; accepted ***}

  \abstract
   {Mid-infrared observations of planet-forming disks reveal a wide diversity in molecular spectra. Carbon and oxygen abundances play a central role in setting the chemical environment of the inner disk and the spectral appearance.}
   {We aim to systematically explore how variations in elemental carbon and oxygen abundances affect the mid-infrared spectra of planet-forming disks, and to identify robust mid-infrared molecular diagnostics of C/H, O/H, and the C/O ratio.}
   {Using the thermochemical disk code \pdm and the line radiative transfer code FLiTs, we construct a grid of 25 models with varying carbon and oxygen abundances, covering a broad range of C/O ratios. We analyze the resulting mid-infrared molecular emission, including species such as \ch{H_2O}, \ch{CO}, \ch{CO_2}, \ch{C_2H_2}, \ch{OH}.}
   {We find that the mid-infrared molecular spectra are highly sensitive not only to the C/O ratio, but also to the absolute abundances of carbon and oxygen. Despite the same disk structure and C/O ratios, molecular fluxes (e.g., \ch{C_2H_2}, \ch{CO_2}) vary by more than an order of magnitude. This variation stems from the differences in excitation conditions and emitting regions caused by the elemental abundances of oxygen and carbon. We identify diagnostic molecular flux ratios - such as \ch{CO_2}/\ch{H_2O} and \ch{H_2O}/\ch{C_2H_2} - that can serve as tracers of C/H and O/H respectively. By combining these diagnostics, we demonstrate a method to infer the underlying C/O ratio.}
   {Our model grid provides a framework for interpreting mid-infrared molecular emission from disks, allowing estimates of elemental abundances if the disk properties and structure are known. Comparisons with recent JWST observations suggest that a variety in C and O abundances is seen in a sample of T Tauri disks, possibly shaped by disk transport processes and the presence of gaps.}


   \maketitle

\section{Introduction}
\label{sec:introduction}
The chemical composition of the inner regions of planet-forming disks influences the resulting bulk composition of the forming planets. Quantitative interpretation of observations requires detailed 2D disk modeling. The inner planet-forming regions, typically at temperatures of a few hundred to a thousand Kelvin, emit in the near- to mid-infrared wavelengths. Infrared observations of T Tauri disks show the presence of \ch{CO}, one of the most abundant and stable molecules in the innermost regions of disks \citep{2003ApJ...589..931N,2007ApJ...655L.105S,2011ApJ...743..112S,2013ApJ...770...94B,2022AJ....163..174B}. The \textit{Spitzer} Space Telescope revealed a rich chemistry within the inner few astronomical units (au) with numerous detections of species such as \ch{H_2O}, \ch{C_2H_2}, \ch{HCN}, \ch{CO_2}, and \ch{OH} at mid-infrared wavelengths \citep{2006ApJ...636L.145L,2008Sci...319.1504C,2011ApJ...733..102C,2008ApJ...676L..49S,2011ApJ...731..130S,2009ApJ...696..143P,2013ApJ...779..178P,2010ApJ...720..887P,2010ApJ...712..274N,2011ApJ...729..145K,2013A&A...551A.118B}. While these observations generally indicated water-rich inner regions for T Tauri disks \citep{2011ApJ...731..130S}, they also revealed a significant chemical diversity, with some sources showing only CO$_2$ and others primarily showing HCN or C$_2$H$_2$ emission \citep{2010ApJ...720..887P}. Crucially, some observed non-detections were likely limited by the sensitivity and spectral resolution of \textit{Spitzer}.

The Mid-Infrared Instrument (MIRI) on board the \textit{James Webb} Space Telescope (JWST) significantly improves upon previous observations with its higher sensitivity and spectral resolution. MIRI reveals a greater diversity of the molecular inventory than was observed with \textit{Spitzer}, detecting fainter molecular features which are important for quantifying the column densities and temperatures of molecules in planet-forming regions \citep[e.g.,][]{2023ApJ...945L...7K,2023ApJ...947L...6G,2023ApJ...957L..22B}. For example, molecules such as \ch{H_2O}, \ch{OH}, \ch{C_2H_2}, \ch{HCN}, as well as \ch{^{13}CO_2}, are now detected in the disk around GW Lup \citep{2023ApJ...947L...6G}, next to the bright \ch{CO_2} emission already seen with \textit{Spitzer}. MIRI also revealed the presence of substantial amounts of warm \ch{H_2O} in a disk with on-going planet formation \citep{2023Natur.620..516P}. The high quality spectra allow for a quantitative analysis of molecular band ratios and thermal gradients within the emitting layers \citep[e.g.,][]{2023A&A...679A.117G,2023ApJ...957L..22B,2024ApJ...962....8S,2024A&A...687A.209K,2024A&A...689A.330T,2024ApJ...975...78R,2025A&A...699A.134T}.

One of the remarkable outcomes of the MIRI observations is the discovery of unexpectedly carbon-rich chemistry in disks around very low-mass stars (VLMS) and brown dwarfs (BD) \citep{2023NatAs...7..805T,2024SciArabhavi,2024A&A...689A.231K,2025ApJ...978L..30L,2025ApJ...986..200F,2025A&A...699A.194A}. While earlier Spitzer data hinted at this \citep{2009ApJ...696..143P,2013ApJ...779..178P}, MIRI has confirmed and expanded upon this finding, revealing an unprecedented diversity and large column densities ($\gtrsim$10$^{20}$ cm$^{-2}$) of carbon-bearing molecules in VLMS/BD disks \citep{2023NatAs...7..805T,2024SciArabhavi}, where \hl{the mid-infrared emission of water is generally weaker than that of carbon-bearing molecules }\citep{2025A&A...699A.194A,2025ApJ...984L..62A}. This contrasts with T Tauri stars, where only one disk with multiple hydrocarbon emission has thus far been identified \citep{2024ApJ...977..173C}.

\hl{The mid-infrared emission from the inner disks observed by MIRI \citep[e.g.,][]{2024PASP..136e4302H,2024ApJ...963..158P} is a complex interplay between the two dimensional disk structure, radiative transfer, excitation conditions, etc.} Tailored modeling studies are therefore the best way to understand how different disk properties and processes affect elemental abundances and mid-infrared spectra in planet-forming disks. Modeling studies show four primary drivers of the mid-infrared spectral appearance of disks: dust properties, disk structure, radiation fields (stellar and interstellar), and elemental abundances \citep[see][]{2009ApJ...704.1471M,2015A&A...582A.105A,2015A&A...582A..88W,2018A&A...618A..57W,2019A&A...631A..81G,2021ApJ...909...55A,2024A&A...683A.219W,2024A&A...682A..91V}. In this paper, we focus on the role of elemental abundances.

Gas-phase elemental abundances, particularly O/H, and C/H ratios, are strongly influenced by radial transport \hl{processes such as pebble drift and advection of gas}, which deliver volatiles such as \ch{H_2O} and \ch{CO_2} to the inner disk modifying the abundances by more than an order of magnitude \citep{2018A&A...611A..80B,2020ApJ...899..134K,2023A&A...677L...7M}. Initially, \hl{inward pebble drift} significantly increases the oxygen abundances in the inner disk, lowering the C/O ratio below solar, a condition that persists briefly ($\lesssim$2 Myr) around VLMS but much longer in T Tauri disks \citep{2023A&A...677L...7M}. The depth of disk gaps critically controls the inward delivery of volatiles \citep{2024A&A...686L..17M}. A shallow gap permits early oxygen-rich material flow, which later shifts to elevated C/O ratios as it accretes onto the star, while a moderately deep gap prolongs the inward oxygen delivery, and a deep gap entirely restricts volatile delivery, depleting the inner disk of oxygen. The location of the gap relative to the ice-lines of major volatiles strongly influences the inner disk composition and its spectral appearance \citep{2021ApJ...921...84K,2023ApJ...954...66K,2025A&A...694A..79S}. Whether such gaps are due to planet formation, or due to internal photoevaporation can determine whether the material from the outer disk reaches the inner disk or is carried away by the photoevaporative winds \citep{2024A&A...691A..72L}. Carbon in dust grains could also be released into the gas phase changing the C/H but not the O/H \citep{2010AdSpR..46...44K,2017ApJ...845...13A,2023NatAs...7..805T,2025A&A...699A.227H}.

Trends in observations have been studied to estimate the elemental abundances, particularly carbon and oxygen, and put them in relation to disk parameters. \citet{2011ApJ...733..102C} and \citet{2013ApJ...766..134N} report a correlation between the observed \ch{HCN}/\ch{H_2O} ratio and the submillimeter disk mass, suggesting a link between the inner disk elemental abundances and planet formation. \citet{2020ApJ...903..124B} expand on this and find that the flux ratios \ch{H_2O}/\ch{HCN}, \ch{H_2O}/\ch{C_2H_2}, and \ch{H_2O}/\ch{CO_2} decrease with an increase in the outer disk dust radii. They also report a possible trend between the stellar luminosity normalized water fluxes and the outer disk radii, which they interpret as an oxygen-enrichment in the inner disk due to inward \hl{pebble} transport. The observed \ch{C_2H_2}/\ch{HCN} flux ratios are less than unity in T Tauri disks and greater than unity in VLMS disks \citep{2009ApJ...696..143P,2025A&A...702A.126G}. Using thermochemical disk models, \citet{2024A&A...689A.231K} show that high column densities of hydrocarbons in VLMS disks require a C/O ratio larger than 1 \citep[see also][]{2026arXiv260123069D}. \citet{2025ApJ...984L..62A} also show that \ch{C_2H_2} fluxes relative to \ch{H_2O} increase with a decrease in stellar luminosities suggesting a higher C/O ratio in sources with lower luminosities \citep[see also][]{2025A&A...702A.126G}. 

While the role of disk properties on the mid-infrared emission has been extensively studied, only a few studies have investigated how mid-infrared molecular fluxes and their ratios vary with changes in elemental abundances. \citet{2011ApJ...743..147N} show that, in the warm surface layers, the column densities of O-bearing molecules such as \ch{H_2O}, \ch{CO_2} and \ch{OH} can decrease and those of C-bearing molecules such as \ch{C_2H_2} and \ch{HCN} can increase by several orders of magnitude when the C/O ratio changes from 0.2 to 4 by varying the oxygen abundance. By varying the carbon abundance, \citet{2018A&A...618A..57W} show that molecular fluxes such as \ch{OH}, \ch{H_2} and ions are relatively stable when the C/O ratio is varied between 0.46 and 1.1, but the fluxes of oxygen-bearing molecules such as \ch{CO_2} and \ch{H_2O} decrease and that of \ch{C_2H_2} increases by up to three orders of magnitude. Similarly, \citet{2021ApJ...909...55A} report variations in molecular flux ratios by changing the initial abundance of \ch{H_2O} in the disk, essentially changing the oxygen budget of the disk (i.e., the C/O ratio \hl{between 0.14 and 0.83}). These modeling studies vary only one elemental abundance at a time $-$ either carbon or oxygen $-$ but \hl{inward material transport by pebble drift and gas advection} can enrich or deplete one or both elements.

In this paper, we systematically explore the effect of changing both the carbon and oxygen elemental abundances on the mid-infrared spectra. In Sect.\,\ref{sec:modelsetupandgrid}, we present the modeling setup and the grid of disk models. The results of the grid are presented in Sect.\,\ref{sec:results}. Reliable molecular diagnostics are presented in Sect.\,\ref{sec:molecular_diagnostics} and put in the context of observations and previous modeling studies in Sect.\,\ref{sec:discussiongrid}. We summarize the conclusions in Sect.\,\ref{sec:conclusions}.

\section{The model setup and grid}
\label{sec:modelsetupandgrid}
We use Protoplanetary Disk Model ({P{\small RO}D{\small I}M{\small O}\footnote{Version 8dedaadb,  \url{https://prodimo.iwf.oeaw.ac.at/}}}, \citealt{woitke2009radiation,kamp2010radiation,thi2011radiation,rab2018x}) to simulate disks with different elemental abundances. The code first initializes the physical structure of the disk such as the gas and dust densities, including dust settling. The code then calculates the dust opacities and solves the continuum radiative transfer, thereby determining the dust temperature structure.  Subsequently, the code computes the chemistry and solves the heating \& cooling balance to find the molecular abundances and the gas temperature structure. We use the Fast Line Tracer (FLiTs, \citealt{2018A&A...618A..57W}) to simulate the mid-infrared spectrum from the resulting \pdm output. \hl{Our models are static models and do not include a treatment of radial transport processes.}

\subsection{The fiducial model}
\label{sec:fiducialmodel}
\hl{We adopt the fiducial T Tauri disk model of \citet{woitke2016consistent} and \citet{2018A&A...618A..57W}, incorporating the smooth inner disk edge introduced by \citet{2024A&A...683A.219W} (also see Appendix \ref{app:OH}).} We adopt the treatment of photorates, molecular shielding, PAHs, dust settling, and escape probabilities from \citet{2024A&A...683A.219W}. The disk model parameters that are common for the entire grid are listed in Table\,\ref{tab:propertiesTable}. The dust opacities are explained in \citet{woitke2016consistent}, and are based on \citet{min2016multiwavelength}. The opacity calculations assume that the grains are spherical and the materials are well mixed. The refractory dust grains are assumed to be composed of 60\% amorphous silicate (\citealt{dorschner1995steps}, {$\rm Mg_{0.7}Fe_{0.3}SiO_3 $}) and 15\% amorphous carbon \citep{zubko1996optical}, with a porosity of 25\%. We use a grain size distribution with grain sizes varying between 0.05\,$\mu$m and 3\,mm with a powerlaw index of -3.5. While the vertically integrated dust size distribution follows a fixed power law, the model calculates the density-dependent settled dust size distribution following \citet{2018A&A...617A.117R}. We do not include viscous heating.

We use steady-state chemistry in all the models presented in this paper. Beyond the large DIANA\footnote{Disk Analysis, \url{https://diana.iwf.oeaw.ac.at/}} chemical network \citep{kamp2017consistent}, we include the extended hydrocarbon chemical network introduced by \citet{2024A&A...681A..22K}. In total, we include 327 species and 13 elements. The elemental abundances of the fiducial model are presented in Table\,\ref{tab:elementsTable}. Similarly to \citet{2018A&A...618A..57W}, we include atomic lines, and the rotational and ro-vibrational molecular transitions in the heating and cooling balance to determine the gas temperatures. The details of the lines included in the mid-infrared wavelength range can be found in Table\,\ref{tab:lineTable}. \hl{Species that have collisional data in the LAMDA database or literature (references in Table\,\ref{tab:lineTable}) are treated in non-LTE. We note that for many polyatomic molecules such data is still missing, which is why we use LTE.}. 

\begin{table}
\caption{Elemental abundances in the fiducial model}
\centering
\begin{tabular}{p{0.13\linewidth}p{0.28\linewidth}p{0.20\linewidth}p{0.20\linewidth}}
\hline \hline
\textbf{Element }  & \textbf{Abundance relative to H ($\epsilon$)} & $\rm \mathbf{12+log_{10}(\epsilon)}$ & \textbf{Mass (amu)}\\
\hline
H   & $1.00\times10^{ 0}$ & 12.00  & 1.0079 \\
He  & $9.64\times10^{-2}$ & 10.984 & 4.0026 \\
C   & $1.38\times10^{-4}$ &  8.14  & 12.011 \\
N   & $7.94\times10^{-5}$ &  7.90  & 14.007 \\
O   & $3.02\times10^{-4}$ &  8.48  & 15.999 \\
Ne  & $8.91\times10^{-5}$ &  7.95  & 20.180 \\
Na  & $2.29\times10^{-9}$ &  3.36  & 22.990 \\
Mg  & $1.07\times10^{-8}$ &  4.03  & 24.305 \\
Si  & $1.74\times10^{-8}$ &  4.24  & 28.086 \\
S   & $1.86\times10^{-7}$ &  5.27  & 32.066 \\
Ar  & $1.20\times10^{-6}$ &  6.08  & 39.948 \\
Fe  & $1.74\times10^{-9}$ &  3.24  & 55.845 \\
PAH & $3.02\times10^{-9}$ &  3.48  & 666.736 \\
\hline
\end{tabular}
\tablefoot{Adopted from \citet{kamp2017consistent}.}
\label{tab:elementsTable}
\end{table}

\subsection{The grid}
\label{sec:thegrid}
\begin{figure}[!ht]
    \centering
    \includegraphics[trim={0 0 0 2em}, clip, width=0.74\linewidth]{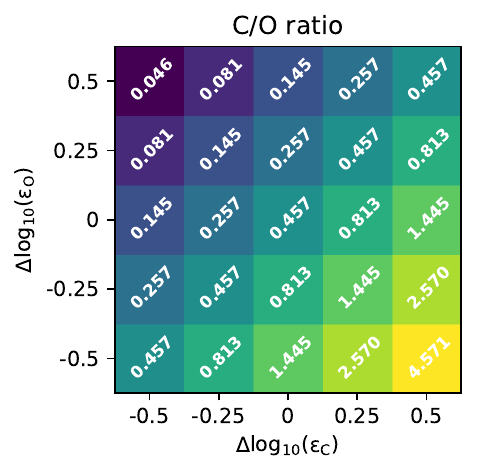}
    \caption{C/O ratios explored in the grid. \hl{The change in the carbon and oxygen elemental abundances relative to their fiducial abundances are shown in the x- and y-axes. The fiducial carbon and oxygen elemental abundances are listed in Table\,\ref{tab:elementsTable}. Each square represents a model in the grid with the C/O ratio written on the square. The color of the square corresponds to the C/O ratio of the corresponding model.}}
    \label{fig:co_grid}
\end{figure}

The objective of this work is to establish reliable molecular diagnostics for carbon and oxygen abundances and their ratios. Hence, we systematically vary their elemental abundances while keeping all remaining parameters fixed to those listed in Table\,\ref{tab:propertiesTable}. \hl{We introduce the IDLI (Infrared Diagnostics from Line emIssion) grid, a uniform set of 25 models constructed by varying the elemental abundances of carbon and oxygen relative to hydrogen ($\epsilon$) by -0.5, -0.25, 0.0, 0.25, and 0.5 dex, spanning a total range of one order of magnitude. This range is physically motivated by dynamic transport models \citep[e.g.,][]{2006Icar..181..178C,2020ApJ...899..134K,2024A&A...686L..17M} and the values explored in previous studies \citep[e.g.,][]{2011ApJ...743..147N,2018A&A...618A..57W,2021ApJ...909...55A}. The parameter space covers C/O ratios higher (up to 4.57) and lower (down to 0.046) than the solar value \citep[0.46,][]{1996ARA&A..34..279S}.} Figure\,\ref{fig:co_grid} shows the carbon-to-oxygen ratio (C/O ratio) of each model in the grid. In the remainder of the paper, we use ($\rm \Delta log_{10}(\epsilon_C)$,$\rm \Delta log_{10}(\epsilon_O)$) to refer to a particular model in the grid. For example, the model with a C/O ratio of 0.046 (top-left in Fig.\,\ref{fig:co_grid}) will be referred to as (-0.5,0.5). All 5x5 panel figures in this paper use the same grid structure, with each panel corresponding to the same carbon and oxygen abundances$-$and C/O ratio$-$as in this figure.

We note that the \hl{physical} processes that change the elemental abundances in the inner disk not only modify the carbon and oxygen abundances, but also the remaining elements. However, exploring the effect of changes in the other elemental abundances is left to a future work. In this work, we vary the elemental abundances in the whole disk and do not consider radial elemental abundance gradients. This approach is appropriate given that mid-infrared emission originates primarily from the innermost disk regions (within a few au), which are typically well-mixed \citep{2022A&A...668A.164W,2006ApJ...647L..57S}.
 
\begin{table}
\caption{Line transitions in the MIRI wavelength range included in the line radiative transfer.}
\centering
\begin{tabular}{p{0.15\linewidth}p{0.20\linewidth}>{\raggedleft\arraybackslash}p{0.10\linewidth}p{0.15\linewidth}}
\hline \hline
\textbf{Species }  & \textbf{Treatment} & \textbf{\#lines} & \textbf{Reference}\\
\hline
\ch{CO}  & non-LTE &  123  & (1) \\
\ch{H_2O}   & non-LTE & 7370  & (2) \\
\ch{OH}  & non-LTE & 219 & (3) \\
\ch{H_2}  & non-LTE &  165  & (4,5) \\
\ch{O}  & non-LTE &  30  & (2,6,7) \\
\ch{CH}  & non-LTE &  4  & (2) \\
\ch{CH_4}  & LTE &  9662  & (8) \\
\ch{C_2H_2}  & LTE &  26457  & (8) \\
\ch{HCN}  & LTE &  5753  & (8) \\
\ch{CO_2}  & LTE &  5746  & (8) \\
\ch{NH_3}  & LTE &  8526  & (8) \\
\ch{SO_2}  & LTE &  8230  & (8) \\
\ch{NO}  & LTE &  4204  & (8) \\
\ch{O_2}  & LTE &  435  & (8) \\
\ch{Ne^+}  & non-LTE &  1  & (9) \\
\ch{Ne^{++}}  & non-LTE &  2  & (2) \\
\ch{Ar^+}  & non-LTE &  1  & (2) \\
\ch{Ar^{++}}  & non-LTE &  3  & (2) \\
\ch{CH^{+}}  & non-LTE &  3  & (2) \\
\ch{Si^{+}}  & non-LTE &  3  & (9) \\
\ch{SH^{+}}  & non-LTE &  4  & (10) \\
\ch{Fe^{+}}  & non-LTE &  10  & (9) \\
\ch{S^{++}}  & non-LTE &  2  & (2) \\
\ch{S}  & non-LTE &  2  & (2) \\
\hline
\multicolumn{2}{c}{Total lines} & 76957 & \\
\hline
\end{tabular}
\tablefoot{
(1) \citet{2013A&A...551A..49T}, (2) LAMDA \citep{2005A&A...432..369S}, (3) See Appendix \ref{app:OH}, (4) \citet{2007MNRAS.382..133W}, (5) \citet{2015MNRAS.453..810L}, (6) \citet{NIST_ASD}, (7) \citet{1996atpc.book.....W}, (8) HITRAN \citep{2022JQSRT.27707949G}, (9) CHIANTI \citep{1997A&AS..125..149D}, (10) John Black (\textit{priv. communication})
}
\label{tab:lineTable}
\end{table}

\section{Results}
\label{sec:results}
\subsection{Fiducial model}
\label{sec:results_fiducial}

\begin{figure*}[!ht]
    \centering
    \includegraphics[width=0.99\linewidth]{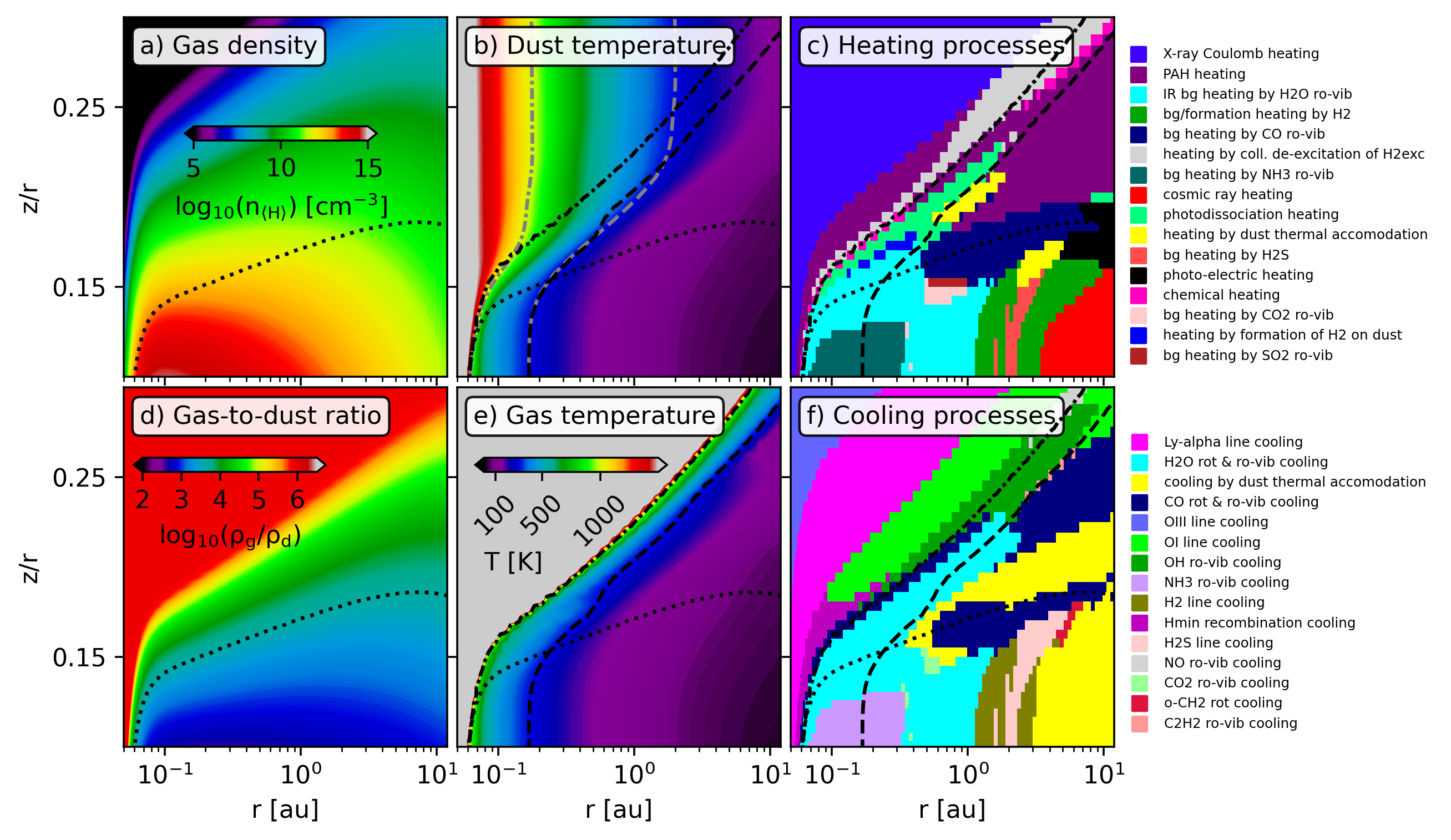}
    \caption{Disk structure of the fiducial model. Only the innermost 12 au and $z/r$$>$0.1 region, which is relevant for MIR emission, is shown. The left panels (a and d) show the gas density and the gas-to-dust mass ratio. The middle panels (b and e) show the dust and the gas temperatures. The right panels (c and f) show the dominant heating and cooling processes, which are labeled to the right of the respective panels. The black dotted line in each panel indicates the vertical extinction A$\rm _v$=1 line. The black dashed and dash-dotted lines show the 300\,K and 1000\,K gas temperature contours (gray contours in panel b indicate dust temperatures).}
    \label{fig:fiducial_summary}
\end{figure*}

\begin{figure*}
    \centering
    \includegraphics[width=0.99\linewidth]{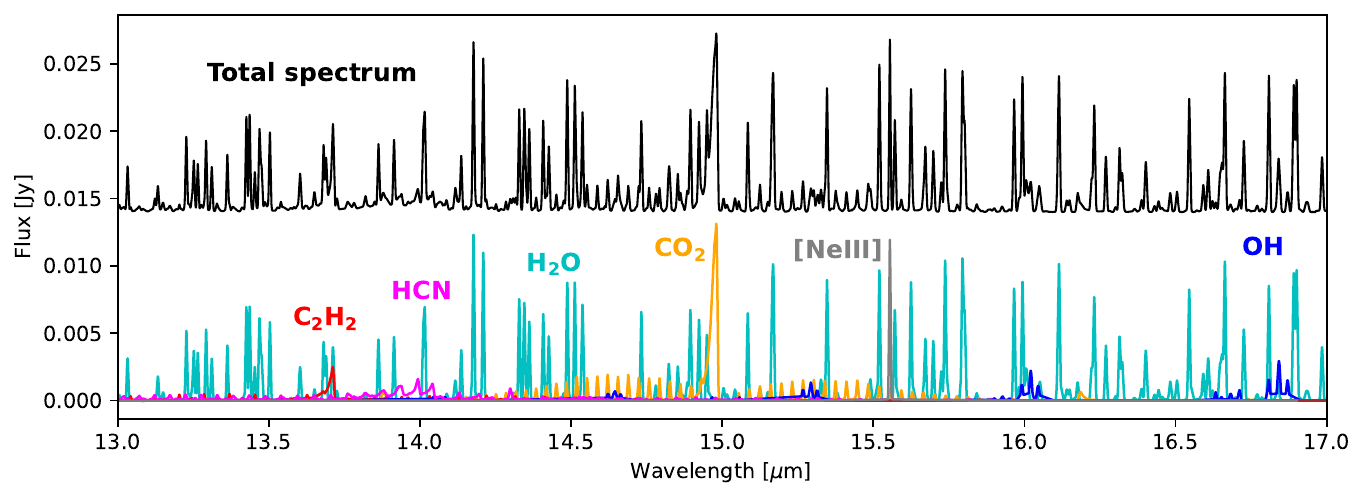}
    \caption{Continuum subtracted FLiTs spectrum (black, shifted vertically for clarity) of reference model using a spectral resolving power of 3000 between 13 and 17\,$\mu$m. The different colors indicate different molecular/atomic emission highlighted in colored text.}
    \label{fig:fiducial_spectrum}
\end{figure*}

Figure \ref{fig:fiducial_summary} provides the summary of the fiducial disk structure (i.e., model (0,0)). Although we assume a total disk gas-to-dust mass ratio of 100, the dust settling leaves the upper regions of the disk with gas-to-dust ratios a few orders of magnitude higher than that. The gas-to-dust mass ratio at the vertical extinction A$\rm _v$=1 layer is about 4200.

The mid-infrared gas emission typically arises from the inner surface regions of the disk ($<$10\,au) with gas temperatures between 1000\,K and 300\,K (Fig.\,\ref{fig:fiducial_summary}). Below the A$\rm _v$=1 layer, the gas and the dust temperatures are well coupled, and above the A$\rm _v$=1 layer, the gas temperatures exceed that of the dust.

In this mid-IR gas emission layer, the dominant processes determining the gas temperature are 1) heating processes such as \ch{H_2O} ro-vibrational heating, heating by UV photodissociation and PAHs (panel c of Fig.\,\ref{fig:fiducial_summary}); 2) cooling processes such as \ch{H_2O} ro-vibrational and rotational cooling, CO rotational and ro-vibrational cooling, and OH ro-vibrational cooling (panel f of Fig.\,\ref{fig:fiducial_summary}). 

The continuum-subtracted mid-infrared spectrum of the model between 13 and 17\,$\mu$m (the molecular-emission-rich part of the spectrum) convolved to a spectral resolving power of 3000 is shown in Fig.\,\ref{fig:fiducial_spectrum}. Clear features of the \ch{CO_2} $Q$-branch as well as the $P$- and $R$-branches are visible. The \ch{C_2H_2} and \ch{HCN} emission are quite weak compared to water and \ch{CO_2}. We find strong emission of pure rotational lines of \ch{H_2O}. The spectrum also shows the \ch{Ne III} atomic fine structure line. We calculate these spectra including all lines in Table\,\ref{tab:lineTable} and a proper treatment of opacity overlap using FLiTs. Although we include \ch{NH_3} and \ch{SO_2} lines in our model, their emission is a few orders of magnitude smaller than the bright emission shown in Fig.\,\ref{fig:fiducial_spectrum}. A detailed analysis of \ch{SO_2} emission can be found in \citet{2025Jelke}. The spectrum of the fiducial model qualitatively agrees with a typical MIRI observations of T Tauri disks, with brighter \ch{H_2O} and \ch{CO_2} features than \ch{C_2H_2} and \ch{HCN}.

\subsection{Properties across the grid}
\label{sec:gridproperties}
The dust temperature, the gas density, and gas-to-dust ratio do not depend on elemental abundances in our models. However, elemental abundances are crucial in calculating the resulting chemistry and the heating-cooling rates, and thus the gas temperatures. The dominant heating and cooling processes change for different C/O ratios in the grid (also shown in Fig.\,\ref{fig:multi_heatcool}). 

When C/O$<$1, photodissociation heating becomes prominent in the warm layers (temperatures between 300\,K and 1000\,K and above A$\rm _v$=1) due to abundant oxygen-bearing molecules such as \ch{H_2O}. The heating by collisional de-excitation of excited \ch{H_2} also dominates. The dominant cooling processes are ro-vibrational and rotational water cooling, OH ro-vibrational cooling, and \ch{[OI]} line cooling. 

When C/O$>$1, the photodissociation heating becomes more prominent than in the case of low C/O ratios. Photodissociation heating is largely due to carbon-bearing molecules. There is also a small contribution of carbon photoionisation heating. On the other hand, the dominant cooling processes are \ch{C_2H_2}, \ch{HCN}, and \ch{CO} ro-vibrational cooling.

\begin{figure*}[!ht]
    \centering
    \includegraphics[width=0.99\linewidth]{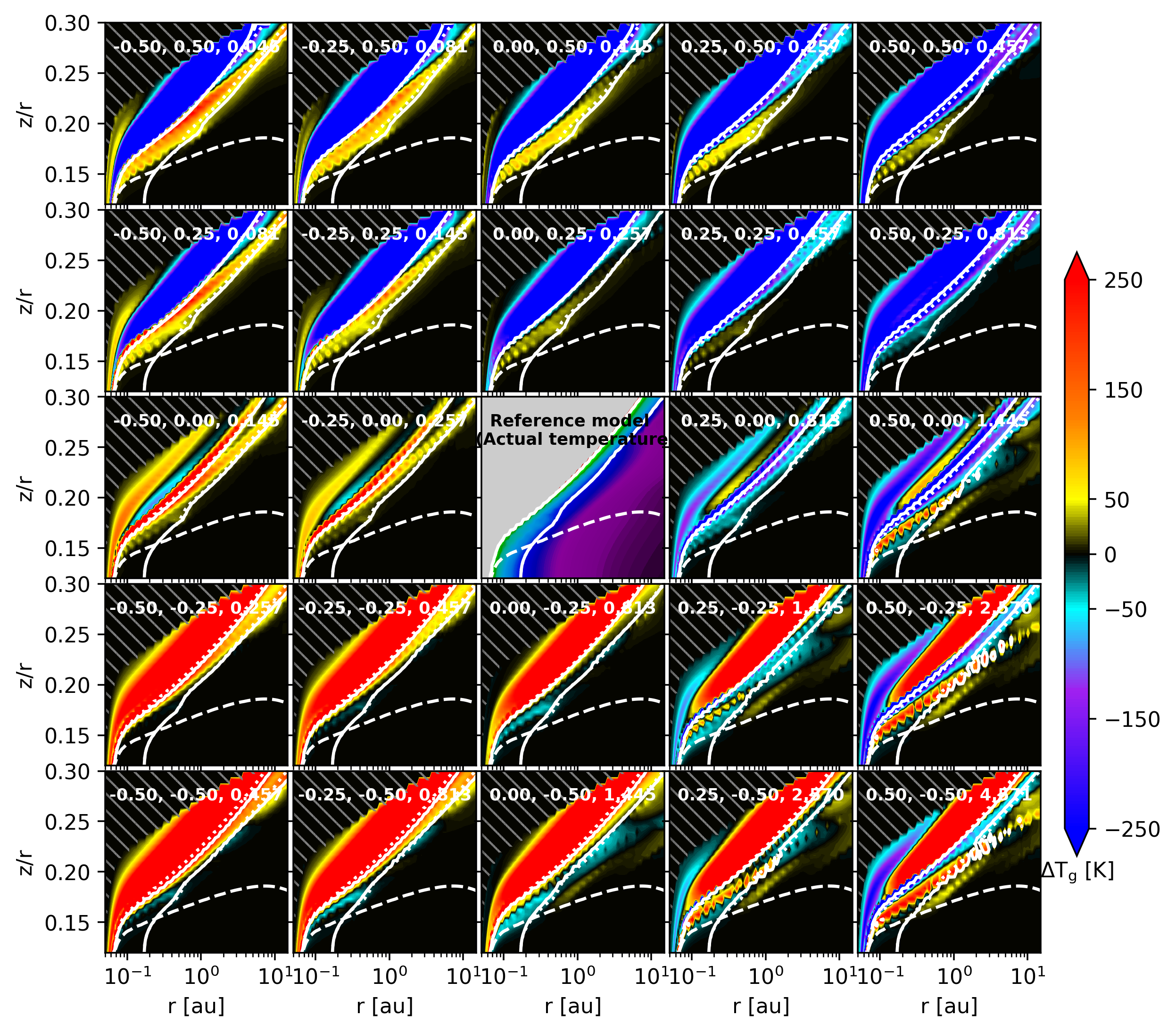}
    \caption{Difference in gas temperatures with respect to the fiducial model. White contours (solid lines) are gas temperatures of 300\,K and 1000\,K. The white dotted lines show the 300\,K and 1000\,K gas temperatures, and \hl{white dashed lines show} A$\rm _v$=1 \hl{of the reference model}. \hl{Since the gas and dust densities are the same for each model, the location of the A$\rm _v$=1 contour is the same in all panels.} The center panel shows the actual gas temperature of the fiducial model, which is the same as panel e of Fig.\,\ref{fig:fiducial_summary}. The text in each panel denotes $\rm \Delta log_{10}(\varepsilon_C)$, $\rm \Delta log_{10}(\varepsilon_O)$, and the C/O ratio.}
    \label{fig:multi_temperature}
\end{figure*}

Figure\,\ref{fig:multi_temperature} shows the differences in gas temperatures between the models in the grid and the reference model. The gas temperatures below the A$\rm _v$=1 layer do not change with respect to the reference model. This is expected because the gas and dust are in thermal equilibrium in these regions as seen in the middle panels of Fig.\,\ref{fig:fiducial_summary}, and hence the elemental abundances have little influence. The figure shows that in general the disk layers above a gas temperature of 1000\,K are much cooler when the oxygen abundances are higher than the reference model (top two rows of panels), while warmer when oxygen is depleted (bottom two rows of panels). This is due to the efficient cooling by neutral oxygen in the models with enhanced oxygen abundances (see Fig.\,\ref{fig:multi_heatcool}).

The diagonal from bottom-left to top-right corresponds to models with the same C/O ratio but different carbon and oxygen elemental abundances. Along these diagonals, the models with higher oxygen content are slightly cooler in the mid-infrared emitting layer (i.e., between the 300\,K-1000\,K contours, above A$\rm _v$=1). This region expands slightly with higher oxygen abundances and shrinks with lower oxygen abundances. This is again due to the dominant cooling processes being \ch{H_2O} and \ch{OH} cooling. The higher the oxygen abundance, the more efficient \ch{H_2O} and \ch{OH} cooling become due to their higher abundances. However, the temperature differences are small ($<$150\,K). \hl{Similarly, increasing the carbon abundance slightly lowers the temperatures in the mid-infrared emitting layer, as the higher CO abundance leads to more efficient ro-vibrational cooling (visible in the top two rows). However, when C/O$>$1, the region below this cooling layer becomes warmer, primarily because the depletion of oxygen reduces the efficiency of \ch{H_2O} cooling.} These results show that the elemental abundances can impact the gas temperatures, which in turn will also affect the mid-infrared spectra.

\section{Molecular diagnostics}
\label{sec:molecular_diagnostics}
In this section, we investigate the resulting mid-infrared spectral appearance across our model grid. Given the established impact of elemental abundances on gas temperatures and chemical abundances, we will specifically examine how molecular fluxes and their ratios can serve as diagnostics for these chemical variations.

\begin{table}[!htbp]
    \centering
    \caption{Wavelength ranges used for calculating the integrated fluxes of molecules.}
    \begin{tabular}{c|c}
       \hline\hline \textbf{Species}  & \textbf{Wavelength range} \\ \hline
       \ch{C_2H_2}$\rm ^a$  & \hl{13.61 - 13.751}\,$\mu$m \\
       \ch{CO_2}$\rm ^a$ & \hl{14.936 - 15.014}\,$\mu$m \\
       \ch{HCN}$\rm ^a$ & \hl{13.837 - 14.075}\,$\mu$m \\
       \ch{OH} & 27.369 - 27.486\,$\mu$m \\
       \ch{CO} & 4.904 - 4.915\,$\mu$m\\
       \ch{H_2O_{rv}} & 6.59 - 6.645\,$\mu$m  \\
       \ch{H_2O_{ro}}$\rm ^a$ & \hl{23.789 - 23.834,  23.878 - 23.912}\,$\mu$m \\
       \ch{NH_3} & 4.9 - 28\,$\mu$m\\ \hline
    \end{tabular}
    \tablefoot{The subscripts `rv' and `ro' refer to ro-vibrational and pure rotational line emission. $\rm ^a$ These wavelength ranges are based on  \citet{2025A&A...694A.147G}.}
    \label{tab:wavelengthintegration}
\end{table}

Table \ref{tab:wavelengthintegration} shows the wavelength ranges used to calculated the integrated fluxes used in the following sections. Using different wavelength regions for the pure rotational lines of \ch{H_2O}, for example 17.19-17.25\,$\mu$m or 14.485-14.562\,$\mu$m, does not change the trends presented in this section.

\subsection{Abundances and line fluxes across the grid}
\label{sec:generalappearance}
The strength of the mid-infrared emission of any species depends on the abundance of the species, the excitation condition, and the dust opacity along the line of sight. As we do not vary the dust properties in our models, their impact on the emission strength of molecular lines is only relevant if the emitting regions themselves vary across models.

\begin{figure*}[!ht]
    \centering
    \includegraphics[width=0.99\linewidth]{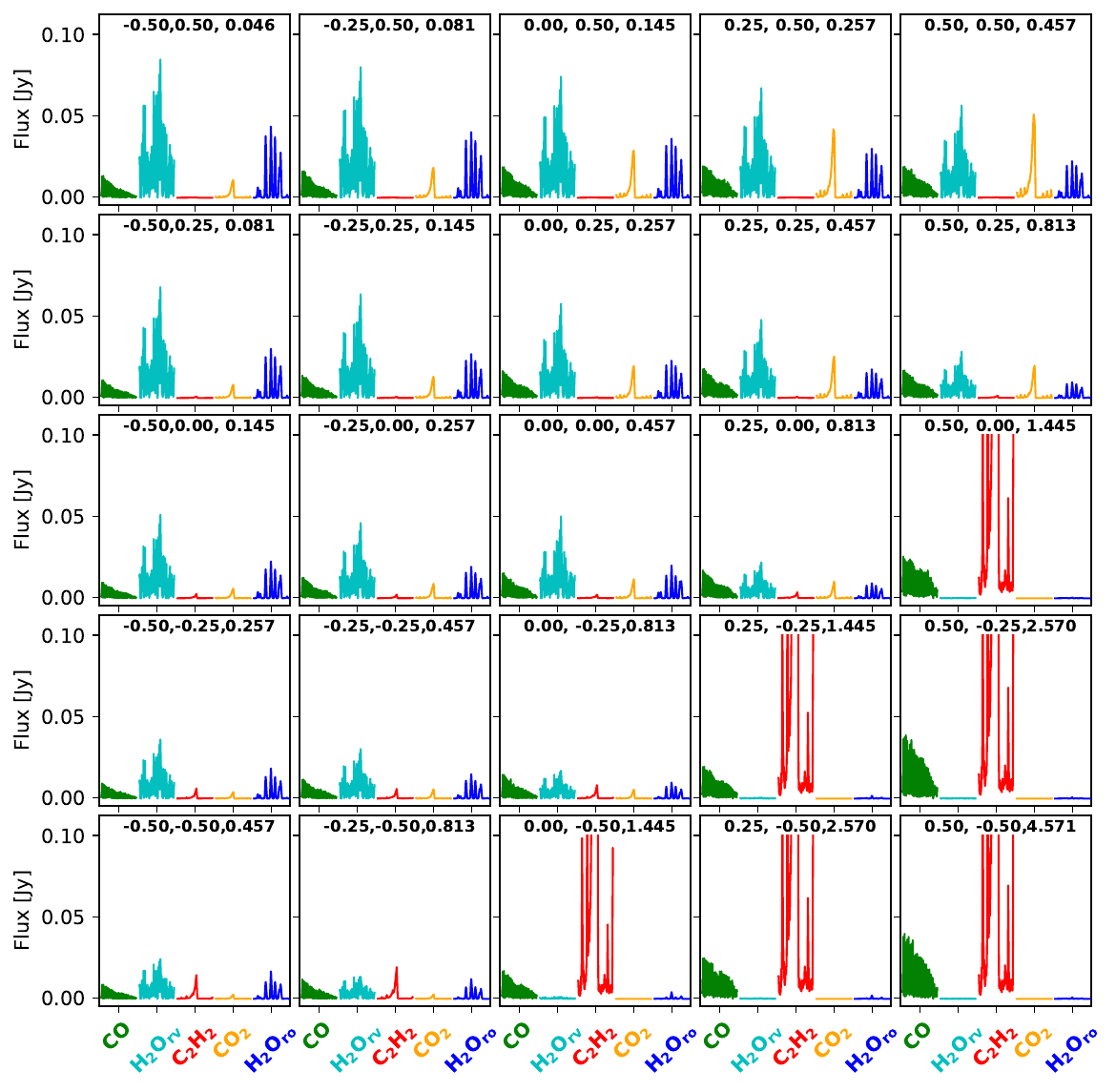}
    \caption{Comparison of molecular emission across the grid. The text in each panel denotes $\rm \Delta log_{10}(\varepsilon_C)$, $\rm \Delta log_{10}(\varepsilon_O)$, and the C/O ratio. The molecular spectra are centered at 5.05\,$\mu$m, 6.6\,$\mu$m, 13.7\,$\mu$m, 14.98\,$\mu$m, and 23.86\,$\mu$m from left to right. Subscripts `rv' and `ro' refer to ro-vibrational and pure rotational line emission.}
    \label{fig:compare_grid_spectra}
\end{figure*}

Figure\,\ref{fig:compare_grid_spectra} provides an overview of the continuum-subtracted spectra of all models in the grid, limited to the wavelength regions corresponding to \hl{emission lines of dominant oxygen and carbon carriers observed in JWST/MIRI observations}$-$the ro-vibrational emission of \ch{CO}, \ch{H_2O}, \ch{C_2H_2}, \ch{CO_2}, and the pure rotational emission of \ch{H_2O}. A quantitative comparison of the integrated fluxes of species listed in Table \ref{tab:wavelengthintegration} across the grid is shown in Fig.\,\ref{fig:gdr_100_fluxsums}. There is a clear diversity in the range of molecular emission strengths across the grid. Moving along the diagonals, i.e., the models with the same C/O ratios, reveals that the molecular strengths change significantly although the C/O ratio remains the same (also see Appendix\,\ref{sec:sameco}). To aid the understanding of the diversity in the spectra presented in Fig.\,\ref{fig:compare_grid_spectra}, it is important to understand the molecular abundances and the line emitting area of different species (see Figs.\,\ref{fig:lineemittingregions_h2o}-\ref{fig:lineemittingregions_co2}). 

\subsubsection{\ch{H_2O}}
The abundance of \ch{H_2O} increases with increasing oxygen abundance, but also increases with decreasing carbon abundance (see Fig.\,\ref{fig:lineemittingregions_h2o}). Correspondingly, the highest water abundance is found in Model (-0.5,0.5) with a C/O ratio of 0.046, and the lowest water abundance is in Model (0.5,-0.5) with a C/O ratio of 4.57. The peak abundance of water in the emitting layers differs by more than three orders of magnitude between these models. Comparing models with $\rm\Delta log_{10}(\epsilon_O)$=0, i.e., for a constant elemental oxygen abundance (middle panels of Fig.\,\ref{fig:lineemittingregions_h2o}), the emitting region of Models (-0.5,0.0), (-0.25,0.0), and (0.0,0.0), have roughly the same water abundances. However, the water abundance in the same surface region in the Model (0.5,0.0) with C/O$>$1 is several orders of magnitude lower, even though the model has the same elemental abundance of oxygen. In this model, oxygen is predominantly in the more stable species \ch{CO}, with \ch{CO} abundances roughly twice those found in Model (0.0,0.0) (Fig.\,\ref{fig:lineemittingregions_co}). Since \ch{CO} is as abundant as \ch{H_2O} in these emitting layers, doubling the \ch{CO} abundance takes away most of the oxygen that would otherwise end up in \ch{H_2O}. Inline with the abundances, the water fluxes (both ro-vibrational and rotational emission) increase with increasing oxygen abundance and decreasing carbon abundance (Fig.\,\ref{fig:compare_grid_spectra}).

\subsubsection{\ch{CO}}
\ch{CO} abundances increase with an increase in carbon or oxygen elemental abundances (Fig.\,\ref{fig:lineemittingregions_co}). However, the emission strength does not necessarily scale with the CO abundance (see Fig.\,\ref{fig:compare_grid_spectra}). This is due to the changes in the emitting regions across the models. 

For models with C/O$<$1, the fluxes increase with increasing carbon or oxygen abundances, but only marginally. In models with the depleted oxygen abundances, the CO emitting region shifts to slightly higher altitudes in the disk (Fig.\,\ref{fig:lineemittingregions_co}). Here, the gas temperatures are higher ($>$1000 K), but the absolute CO abundances are lower. Conversely, in models with enhanced oxygen, CO emission originates from regions of lower temperature ($<$1000 K) but higher absolute abundances.

In the models with C/O$>$1, particularly the models with enhanced carbon (by 0.5 dex) have much larger emitting areas that probe cooler temperatures. Although neither \ch{CO} abundances nor the temperature within the emitting regions of these models are the highest in the grid. However, the emitting area is more than five times larger than in models with C/O$<$1, this leads to the strongest CO fluxes in the grid.

\subsubsection{\ch{C_2H_2}}
One of the most prominent trends in the simulated spectra is the bright emission of \ch{C_2H_2} in models with C/O>1 (Fig.\,\ref{fig:compare_grid_spectra}). In these high C/O models, we find large abundances of \ch{C_2H_2} that form in the disk surface layers (Fig.\,\ref{fig:lineemittingregions_c2h2}). While in the rest of the models, the emitting region is limited to the innermost regions close to the rim, the C/O$>$1 models have emitting areas several orders of magnitude larger.

In models with C/O$<$1, the \ch{C_2H_2} emission is generally weaker. This is largely because the bulk of the \ch{C_2H_2} resides below the $\tau$=1 layer, leaving little \ch{C_2H_2} that can emit at mid-infrared wavelengths (see Fig.\,\ref{fig:lineemittingregions_c2h2} and \ref{fig:lineemittingregions_c2h2_zoom}). In these models, we find that the emission strength of \ch{C_2H_2} is more strongly influenced by the abundance of oxygen compared to the abundance of carbon. For example, in the models with C/O=0.457 (diagonal panels), we see a diversity in the \ch{C_2H_2} emission strengths (Fig.\,\ref{fig:compare_grid_spectra} and \ref{fig:co_change}). Although Model (0.5,0.5) has an order of magnitude higher carbon abundance than Model (-0.5,-0.5), the \ch{C_2H_2} emission is more than an order of magnitude brighter in the latter. This is because oxygen-depletion leads to an increase in \ch{C_2H_2} abundance (also see \citealt{2026Esteve}). The \ch{C_2H_2} abundance in the innermost region above the $\tau$=1 line increases by several orders of magnitude when $\rm \Delta log_{10}(\epsilon_O)$ decreases from +0.5 to -0.5 (top panels to bottom panels Fig.\,\ref{fig:lineemittingregions_c2h2_zoom}). In addition, the abundance of \ch{C_2H_2} also increases in the thin radially extended surface layer which also contributes to the observed fluxes. However, the contribution to the total flux from this low density surface layer is less than 33\,\% even though the emitting area is several orders of magnitude larger. It is not clear whether \ch{C_2H_2} excitation is in local thermodynamic equilibrium (LTE) in this surface layer since densities are of the order of $\geq$10$^{10}$\,cm$^{-3}$. Typical critical densities for these ro-vibrational lines are $\gtrsim$10$^{12}$\,cm$^{-3}$. However, infrared pumping can also play a role, possibly bringing level populations again closer to LTE.

\subsubsection{\ch{CO_2}}
\ch{CO_2} is the most strongly affected molecule by both carbon and oxygen abundances (Fig.\,\ref{fig:compare_grid_spectra}). In the models with C/O$<$1, the \ch{CO_2} emission strength varies by a factor of $\sim$30. More interestingly, this variation occurs among models of the same C/O, i.e. for models with C/O=0.457.

The abundance of \ch{CO_2} increases with an increase in the elemental carbon and oxygen abundances (Fig.\,\ref{fig:lineemittingregions_co2}). On the other hand, the emitting region moves to a cooler region with a decrease in the elemental carbon and oxygen abundances. Both of these together explain the large change in the emission strength across the grid.

In the models with C/O$>$1, the \ch{CO_2} abundances are inherently lower, and the reservoir largely resides below the $\tau$=1 layer.

\subsubsection{OH, HCN, NeII, NH$_3$}
Fluxes of \ch{OH}, \ch{HCN}, \ch{[Ne II]}, and \ch{NH_3} also show some trend (Fig.\,\ref{fig:gdr_100_fluxsums}). Similar to \ch{H_2O}, \ch{OH} also peaks when carbon is depleted and oxygen is enhanced. However, the drop in \ch{OH} flux for models with C/O$>$1 is not as drastic as in \ch{H_2O} flux, due to the larger emitting area of \ch{OH} (see Appendix\,\ref{sec:miscmolecules}). It is important to note that while our models include an approximate treatment of non-LTE, other processes such as chemical excitation (i.e. prompt emission of \ch{OH}, \citealt{2021A&A...650A.192T}) are not included. This latter excitation mechanism can be very important for predicting accurate emission strengths of \ch{OH} in disks (e.g. \citealt{2024A&A...691A..11T}), but less important for a comparative study between models. \ch{HCN} fluxes behave similar to \ch{C_2H_2} fluxes, showing high fluxes when C/O$>$1. The \ch{[Ne II]} emission depends on the oxygen abundance, and is only slightly influenced by the carbon abundance. However, the variation is small compared to the other species studied here. The dependence on the oxygen abundance is largely related to the dominant cooling processes in the emitting region of \ch{[Ne II]} (see Appendix\,\ref{sec:miscmolecules}). \ch{NH_3} is brighter when the oxygen abundance is higher and carbon abundance is lower. We note that the peak of the brightest \ch{NH_3} emission in our models is about only 0.2\,mJy, largely due to \ch{NH_3} residing closer to the midplane ($z/r$$<$0.15, see Appendix\,\ref{sec:miscmolecules}).

\begin{figure*}[!ht]
    \centering
    \includegraphics[width=0.99\linewidth]{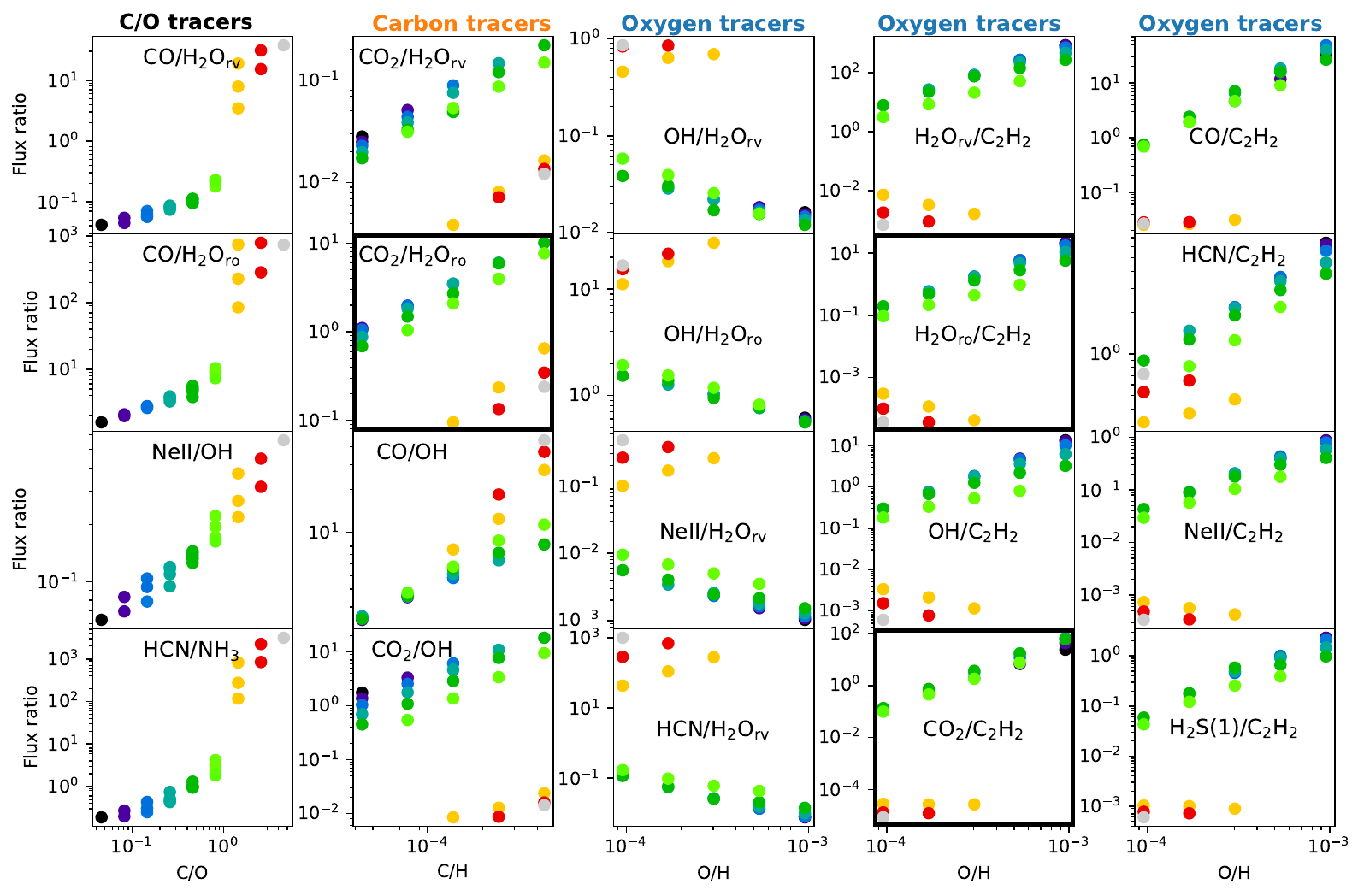}
    \caption{Mid-infrared molecular diagnostics: integrated flux ratios. The colors of the scatter points indicate the C/O ratio. \hl{All species pairs in our models which show trends are shown.} Reliable tracers are highlighted with thick panel edges.}
    \label{fig:gdr_100_fluxratios_CO_diagnostics}
\end{figure*}

\subsubsection{Summary}
Changes in molecular abundances and the properties of the emitting region, such as temperature and spatial extent, are influenced by elemental abundances, leading to variations in the molecular line fluxes. For example, \ch{C_2H_2}, which contains no oxygen, is still strongly influenced by the oxygen abundance. Conversely, \ch{H_2O}, which lacks carbon, is affected by both carbon and oxygen abundances. Molecules like \ch{CO} and \ch{CO_2}, which include both carbon and oxygen, are influenced by the abundances of both elements, though the effect is more pronounced for \ch{CO_2}. In models where the C/O ratio exceeds 1, the chemistry changes significantly compared to models with C/O$<$1, and becomes dominated by hydrocarbons such as \ch{C_2H_2} and depleted in \ch{H_2O}, even when oxygen is not significantly depleted.

\subsection{\hl{Line flux ratios as tracers} of C/O, C/H, and O/H}
Figure\,\ref{fig:compare_grid_spectra} shows the diversity in the emission strengths of different molecules with the elemental abundances of carbon and oxygen. The figure also shows that for the same C/O ratio, the molecular fluxes (absolute and relative) can be significantly different.

\hl{In the following subsections, we present mid-infrared diagnostics for carbon and oxygen abundances based on line flux ratios. While we explore ratios involving all species in the model spectra, we mostly focus here on a few species. Emission from atoms and ions arise from the thin tenuous layers that do not necessarily reflect the bulk composition of the disk, often affected by factors such as disk winds and jets. Molecules such as CO and OH are expected to be excited by non-LTE processes such as UV and IR pumping, chemical pumping, beyond collisional excitation. For example, \citet{2024A&A...686A.117T} demonstrate that spectral resolving power higher than that of JWST/MIRI is required to resolve CO lines to uncover the broad and narrow components of CO emission, the latter of which is potentially linked to disk winds. \citet{2024A&A...686A.117T} also show that the broad component should arise from within the dust sublimation radius, whereas the bulk of the emission from other molecules arises radially further out in the disk. However, other polyatomic molecules such as \ch{CO_2}, \ch{C_2H_2}, \ch{H_2O} (rotational lines), and \ch{HCN}, which are commonly detected in the mid-infrared spectra of T Tauri disks, have been reproduced sufficiently well with LTE approximation across a wide range of temperatures, column densities, and emitting radii \citep[e.g.,][]{2023ApJ...947L...6G,2023A&A...679A.117G,2023NatAs...7..805T,2023ApJ...957L..22B,2023ApJ...959L..25X,2024ApJ...975...78R,2024A&A...687A.209K,2024A&A...689A.330T,2025A&A...699A.134T,2025A&A...693A.278V}. So we focus mainly on these molecules \hll{(e.g., \ch{CO_2}, \ch{C_2H_2}, \ch{H_2O})}.}

\subsubsection{Tracers of C/O}

As presented earlier, the emission of different species react differently to the C/O ratio in the disk. We compared the C/O ratios of all models with the flux ratios between all species in Table.\,\ref{tab:wavelengthintegration}, as well as [NeII] and \ch{H_2} S(1), to identify any interesting species pair whose flux ratio can be used to estimate the C/O ratio. We found four flux ratios \hl{showing some trend with C/O ratio}: \ch{CO}/\ch{H_2O_{rv}}, \ch{CO}/\ch{H_2O_{ro}}, \ch{[Ne II]}/\ch{OH}, and \ch{HCN}/\ch{NH_3}, as shown in the left panels of Fig.\,\ref{fig:gdr_100_fluxratios_CO_diagnostics}.

Except for the flux ratio \ch{[Ne II]}/\ch{OH}, the remaining three flux ratios increase substantially when the C/O ratios are greater than unity. Below C/O ratios of unity, the flux ratios show a positive trend with the C/O ratio. For a given C/O ratio, i.e., corresponding to models with different oxygen and carbon abundances, the flux ratios show little spread. However, beyond the C/O ratio of unity, the spread increases and the trend is not quite evident. 

These flux ratios are unreliable to be applied to observations. Ro-vibrational band emission of \ch{H_2O} and \ch{CO}, and \ch{OH} are influenced by non-LTE effects, and the flux levels of \ch{NH_3} are below the noise levels of MIRI. So far \ch{NH_3} has not been observed in emission in the mid-infrared wavelengths in any planet-forming disks. The strength of the \ch{[Ne II]} emission observed is often influenced by jets and disk winds, and by the X-ray luminosity \citep{2010A&A...519A.113G,2020ApJ...903...78P}. 

\subsubsection{Tracers of C/H}
\label{sec:c/h}
Here we explore tracers for measuring the C/H and O/H individually such that by combining these tracers the C/O can be determined. Hence, we now compare the C/H ratios (or the elemental carbon abundances) of all models of the grid with the flux ratios between the species. We find four tracers: \ch{CO_2}/\ch{H_2O_{rv}}, \ch{CO_2}/\ch{H_2O_{ro}}, \ch{CO}/\ch{OH}, and \ch{CO_2}/\ch{OH} (Fig.\,\ref{fig:gdr_100_fluxratios_CO_diagnostics}). Except for the flux ratio \ch{CO}/\ch{OH}, the remaining three flux ratios are substantially smaller in models with C/O$>$1 compared to models with C/O$<$1 for the same carbon abundances. However, in both cases, the flux ratios show a positive trend with the carbon abundance with a small spread. Again, discarding the flux ratios involving the ro-vibrational band emission of \ch{H_2O} and \ch{CO}, and \ch{OH}, we propose the flux ratio between \ch{CO_2} and pure rotational lines of \ch{H_2O} to be reliable C/H tracers. \hl{We note that the two distinct trends of the \ch{CO_2}/\ch{H_2O} ratio (for models with C/O$<$1 and C/O$>$1) could theoretically result in degenerate solutions for C/H if abundances slightly higher or lower than our grid values were considered. However, such degeneracies are lifted when complementary flux ratios are included in the analysis (see Sect.\,\ref{sec:robustcotracer}).}

\begin{figure*}[!ht]
    \centering
    \includegraphics[trim={0 1.5cm 0 0},clip,width=0.89\linewidth]{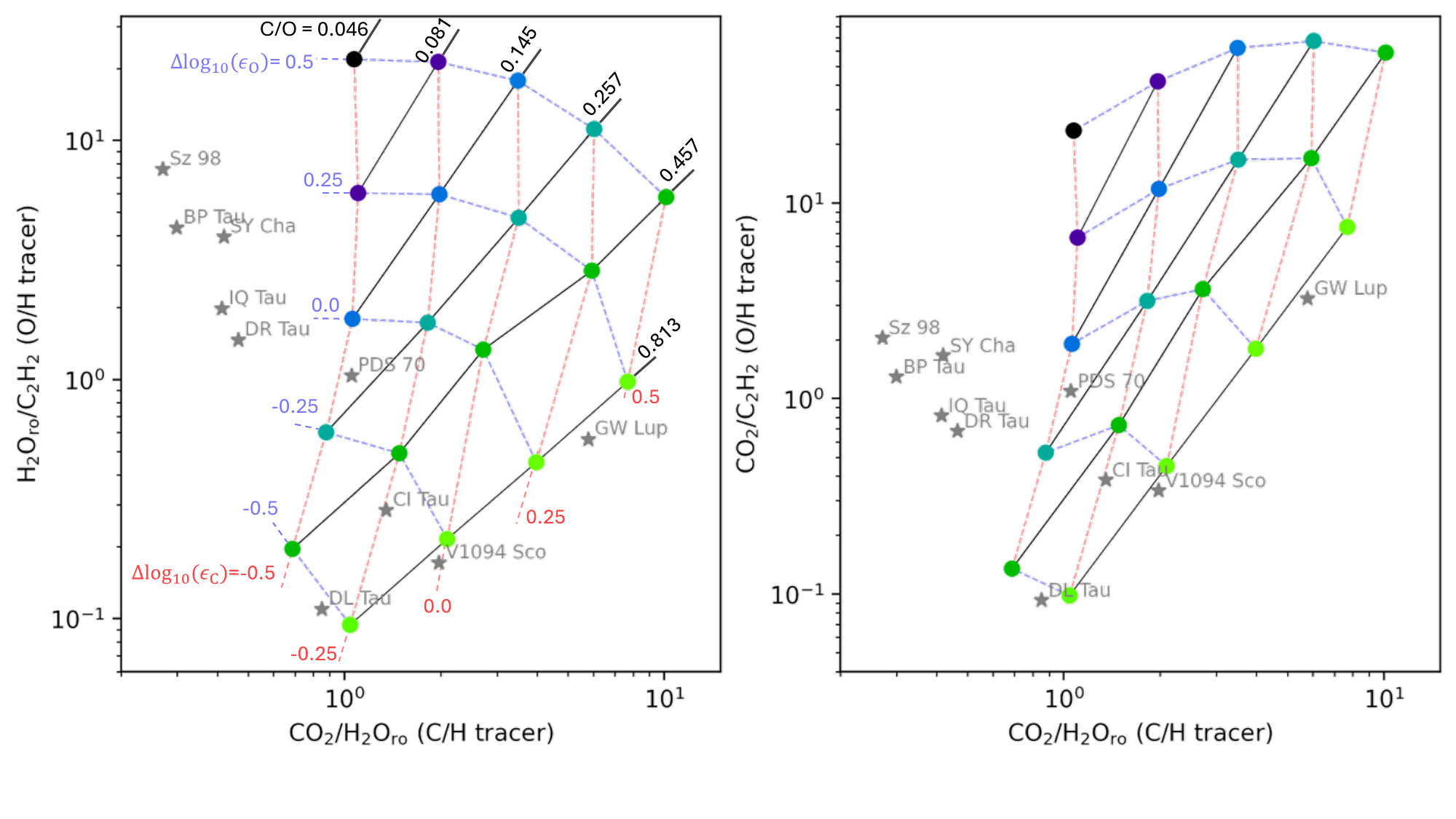}
    \caption{Proposed C/O diagnostics: integrated flux ratios \ch{H_2O}/\ch{C_2H_2} vs \ch{CO_2}/\ch{H_2O} (left) and \ch{CO_2}/\ch{C_2H_2} vs \ch{CO_2}/\ch{H_2O} (right). The colors of the scatter points indicate the C/O ratio (see Fig.\ref{fig:gdr_100_fluxratios_CO_diagnostics}). C/O ratio contours are shown as black lines, O/H contours as cyan lines, and C/H contours as red lines. The grey stars indicate the observations. The observed fluxes are taken from \citet{2025A&A...694A.147G}.}
    \label{fig:obs_c_o_tracer}
\end{figure*}

\subsubsection{Tracers of O/H}
\label{sec:o/h}
Figure\,\ref{fig:gdr_100_fluxratios_CO_diagnostics} shows several tracers of the elemental oxygen abundance O/H. Again, the flux ratios are substantially different when the C/O ratios are larger than unity. The middle panels of figure corresponding to \ch{OH}/\ch{H_2O_{rv}}, \ch{OH}/\ch{H_2O_{ro}}, \ch{[Ne II]}/\ch{H_2O_{rv}}, and \ch{HCN}/\ch{H_2O_{rv}} show a negative trend with O/H with a small spread when the C/O ratios are smaller than unity. However, each of these flux ratios involve either \ch{OH}, \ch{[Ne II]}, or the ro-vibrational band emission of \ch{H_2O}.

The remaining oxygen tracers are the flux ratios with \ch{C_2H_2}. This is not surprising, as discussed in Sect.\,\ref{sec:generalappearance} the \ch{C_2H_2} emission is strongly influenced by the abundance of oxygen (i.e., O/H). Among these tracers, the flux ratios of the pure rotational lines of \ch{H_2O} and \ch{CO_2} with \ch{C_2H_2} are reliable. Moreover, the flux ratio \ch{CO_2}/\ch{C_2H_2} has the least spread for a given oxygen abundance.

\subsection{Robust diagnostics for measuring C/O}
\label{sec:robustcotracer}
In Sects.\,\ref{sec:c/h} and \ref{sec:o/h} we identified reliable tracers of elemental abundances of oxygen and carbon. Since our grid of models are equally spanning the oxygen and carbon abundances as shown in Fig.\,\ref{fig:co_grid}, we expect to reconstruct the matrix in Fig.\,\ref{fig:co_grid} using these tracers identified.

In Fig.\,\ref{fig:obs_c_o_tracer} we plot the diagnostic flux ratios with the oxygen tracers on y-axis against the carbon tracer on x-axis: \ch{H_2O}/\ch{C_2H_2} vs \ch{CO_2}/\ch{H_2O} (left panel), and \ch{CO_2}/\ch{C_2H_2} vs \ch{CO_2}/\ch{H_2O} (right panel). Models with C/O$>$1 are not shown in either panels as the oxygen tracer flux ratios of these models lie \hl{more than two} orders of magnitude lower than the rest (as shown in Fig.\,\ref{fig:gdr_100_fluxratios_CO_diagnostics} and \ref{fig:obs_c_o_tracer_co_1}). The ideal case for elemental abundance retrieval would be for the diagnostic grid (Fig.\,\ref{fig:obs_c_o_tracer}) to reproduce the rectangular pattern of abundance variations shown in Fig.\,\ref{fig:co_grid}. However, we find a slightly skewed but still systematically organized grid using these diagnostic tracers. In both panels, both axes span more than an order of magnitude. The black lines, which represent constant C/O ratios, show a relatively clean progression across both panels. This indicates that the combination of these flux ratios is effective in diagnosing the C/O ratio. The cyan and red lines (indicating the constant O/H and C/H values) are somewhat tilted or skewed, but still follow an orderly trend. While the C/H and O/H could not be directly retrieved from the individual flux ratios due to the spread between models of different C/O ratios as shown in Fig.\,\ref{fig:gdr_100_fluxratios_CO_diagnostics}, using the combination of their flux ratios allows distinguishing the elemental abundances. 

\section{Discussion}
\label{sec:discussiongrid}
\subsection{Grid in the context of the observations}
\label{sec:context_observations}
In Sect.\,\ref{sec:generalappearance} we established that flux ratios that trace the oxygen and carbon abundances can be combined to predict C/O ratios in our models. In this section, we compare the flux ratios from our models to that measured by observations in the mid-infrared. Our models pertain to a single representative disk. In reality, disk and stellar properties differ from source to source. This can significantly influence the fluxes and flux ratios measured. However, the \hl{line ratio} diagnostics identified in the previous section span more than an order of magnitude. Even if the model properties do not necessarily reflect the actual source properties, we expect the general trend predicted by our models to be robust.

Figure\,\ref{fig:obs_c_o_tracer} shows the flux ratios calculated from the fluxes reported by \citet{2025A&A...694A.147G} for 10 T Tauri disks on top of the model results. \citet{2025A&A...694A.147G} only include sources with confirmed detections of gaps in millimeter dust. The advantage of using this sample in our comparison is that dynamic disk models indicate that disk gaps can affect the elemental abundances in the inner disk \citep{2021ApJ...921...84K,2023ApJ...954...66K,2024A&A...691A..72L,2024A&A...686L..17M,2025A&A...694A..79S}. Since the mid-infrared fluxes in our models are not affected by the disk geometry beyond a few astronomical units (au), and the smallest gap reported in the sample is about 9\,au, we essentially compare the effect of the varying elemental abundances on the flux ratios. Some caveats are discussed later.

Figure\,\ref{fig:obs_c_o_tracer} shows that the range of observed flux ratios that trace the oxygen abundance \hl{(\ch{CO_2}/\ch{C_2H_2} and \ch{H_2O_{ro}}/\ch{C_2H_2})} spans roughly the same range as our models, but the observed flux ratios that trace the carbon abundance \hl{(\ch{CO_2}/\ch{H_2O})} extend a factor 2 lower than the models with C/O$<$1. This could largely be due to the specific disk properties of these sources that could be different from those assumed in our models. However, the range of flux ratios show important evidence for difference in elemental abundances in these disks. Sources such as Sz\,98, BP\,Tau and SY\,Cha show high \ch{H_2O}/\ch{C_2H_2} flux ratio indicating elevated abundances of oxygen. IQ\,Tau, and DR\,Tau correspond roughly to solar oxygen abundance. CI\,Tau, V1094\,Sco, and DL\,Tau indicate depleted oxygen abundances. CI\,Tau, DL\,Tau, V1094\,Sco, and GW\,Lup have relatively bright \ch{CO_2} emission indicating higher carbon abundances than the oxygen-rich sources. With the highest \ch{CO_2}/\ch{H_2O} flux ratio and moderately high \ch{H_2O}/\ch{C_2H_2} ratio, the inner disk of GW\,Lup could have enhancements in both carbon and oxygen.

While deep gaps completely block the inward transport of material, leaky gaps allow rapid inward transport; both scenarios lead to oxygen depletion in the inner disk \citep{2021ApJ...921...84K,2024A&A...686L..17M}. In contrast, a moderately leaky gap can prolong the transport process, helping to maintain an oxygen-rich inner disk \citep{2025A&A...694A.147G}. The sources in the top-left part of Fig.\,\ref{fig:obs_c_o_tracer} (left panel) are indeed expected to have either moderately leaky gaps or deep gaps where gas has been detected in the gap indicating inward material transport. IQ\,Tau and DR\,Tau have similar disk structure \citep{2025A&A...694A.147G} and are clustered together in the figure. CI\,Tau is expected to have a deep gap, while V1094\,Sco and DL\,Tau are expected to be either old disks with moderate gaps or young disks with deep gaps, in either case depleted in oxygen. Interestingly, no source in this sample resides in the top-right part of the figure, indicating equally enhanced carbon and oxygen abundances. GW\,Lup could, however, still be an example of such enhancement and would require a dedicated thermo-chemical disk model to examine the abundances more accurately. No source has \ch{H_2O}/\ch{C_2H_2} or \ch{CO_2}/\ch{C_2H_2} flux ratios \hl{(O/H-tracers)} far below the region shown in the figure, where our models with C/O$>$1 lie (see Fig.\,\ref{fig:obs_c_o_tracer_co_1}).

A comparison of observations with the model fluxes similar to Fig.\,\ref{fig:obs_c_o_tracer}, but with observed flux ratios of disks around very low-mass stars is provided in Appendix\,\ref{sec:vlms_obs}.

\subsection{Comparison to previous modeling studies}
\citet{2021ApJ...909...55A} explore the effects of elemental abundances on mid-infrared fluxes. They change the abundance of \ch{H_2O} to achieve a change in the C/O ratio, essentially changing the oxygen abundances only. According to our analysis, such variation should lead to large changes in molecular flux ratios of the oxygen tracers \hl{(\ch{H_2O_{ro}}/\ch{C_2H_2} and \ch{CO_2}/\ch{C_2H_2})} but not so for the carbon tracer \hl{(\ch{CO_2}/\ch{H_2O})}. And indeed, they report large variations in the flux ratios (up to three orders of magnitude) of \ch{CO_2}/\ch{C_2H_2} and \ch{H_2O}/\ch{C_2H_2}, but very little variation in the \ch{CO_2}/\ch{H_2O} ratio. However, the specific values of these variations do not match between the two studies. This could be due to differences in the modeling approaches. For example, we perform a full line radiative transfer solution to predict the model spectra, but \citet{2021ApJ...909...55A} use LTE slab models to predict the spectra considering the column of molecules down to a depth where the total \ch{H_2} gas column densities reach 1.8$\times$10$^{23}$\,cm$^{-2}$ and up to 10\,au in radius. We clearly see that for molecules such as \ch{H_2O}, \ch{CO}, and \ch{CO_2}, only a thin layer in the disk surface that is radially limited contributes to the total flux.

\citet{2018A&A...618A..57W} show the sensitivity of the fluxes close to a C/O ratio of unity \hl{by varying the elemental carbon abundance}. They report a decrease in flux of \ch{CO_2} and an increase in the flux of \ch{C_2H_2} by several orders of magnitude when pushing models to values of C/O$>$1. This is in line with our models. However, the relative change in flux of the individual molecules such as \ch{H_2O}, or \ch{CO_2} in our models are not as drastic as reported by \citet{2018A&A...618A..57W}. This difference could stem from several factors including slightly different disk structures, and different treatment of escape probabilities and shielding \citep[see][]{2024A&A...683A.219W}. Moreover, \citet{2018A&A...618A..57W} report integrated fluxes of all lines of a molecule.

\subsection{Caveats}
\hl{While observed oxygen tracers align well with our C/O$<$1 models, the observations do not overlap with the model grid when oxygen and carbon tracers are considered simultaneously}. One of the important factors that could affect the \ch{CO_2}/\ch{H_2O} ratio is the inner disk radius. \citet{2018A&A...618A..57W}, \citet{2021ApJ...909...55A} and \citet{2024A&A...682A..91V} show that the inner disk radius has a distinct effect on the mid-infrared fluxes. The fluxes increase with an increase in inner radius due to the larger emitting area of the inner rim, but only up to about 5\,au, beyond which the fluxes drop due to lower rim temperatures (lines cannot be excited anymore). The fluxes can vary by more than an order of magnitude and this effect depends on the molecule, thus strongly influencing the \ch{CO_2}/\ch{H_2O} flux ratio. For example, BP Tau is expected to have a close-in gap or a large cavity; the latter could strongly boost the \ch{CO_2}/\ch{H_2O} flux ratio making it appear water-rich. \citet{2024A&A...682A..91V} show that variations in the \ch{CO_2}/\ch{H_2O} molecular flux ratio, driven by changes in the inner disk radius, can exceed an order of magnitude — comparable to the range shown in Fig.\,\ref{fig:obs_c_o_tracer}. \citet{2022A&A...668A.164W} show that with the same elemental abundances vertical mixing can drive an active organic chemistry, thus enhancing hydrocarbon fluxes. Variability in inner disk turbulence between sources can cause further spread in range of elemental abundances that reproduce the same flux ratios.

Additionally, previous modeling studies have shown that dust properties and disk structures can play a large role in the spectral appearance of the disk, since they determine the amount and shape of radiation field that enters or leaves the disk. For example, the gas-to-dust ratio, maximum dust grain size, dust settling, and the dust size distribution can vary the mid-IR line strengths by more than an order of magnitude \citep{2009ApJ...704.1471M,2015A&A...582A.105A,2018A&A...618A..57W,2019A&A...631A..81G,2024A&A...683A.219W}. Further, in this study, we assume fixed stellar properties. However, the observed sources span stellar luminosities between 0.22-1.52\,$L_{\odot}$. The mass and luminosity of the central object can strongly influence the molecular abundances \citep{2015A&A...582A..88W}. A stronger ultraviolet radiation field (from the star or the interstellar medium) or the presence of X-rays greatly enhances the mid-infrared fluxes \citep{2015A&A...582A.105A,2015A&A...582A..88W, 2026A&A...705A.222K}.

Although multiple additional factors can influence the predicted fluxes and flux ratios, the change of elemental abundances in our grid largely explains the spectral diversity of the observations. More detailed analysis is required, that takes into account the source specific properties, to constrain the elemental abundances in individual sources.

\section{Conclusions}
\label{sec:conclusions}
We have constructed a grid of thermochemical disk models that vary both the carbon and oxygen elemental abundances, allowing for a systematic exploration of their combined effects on mid-infrared molecular emission. Our key findings are as follows:
\begin{enumerate}
    \item Non-uniqueness of the C/O ratio: Molecular fluxes and ratios cannot be solely attributed to the C/O ratio; models with identical C/O ratios but different absolute C and O abundances can yield significantly different spectra.
    \item Sensitivity of individual species: Molecules such as \ch{H_2O}, \ch{C_2H_2}, and \ch{CO_2} are particularly sensitive to the elemental abundance of carbon and/or oxygen. Notably, \ch{C_2H_2}, though a purely carbon-bearing molecule, is strongly affected by oxygen depletion, while \ch{H_2O} and \ch{CO_2} is influenced by both elements.
    \item Diagnostic tracers: We identify flux ratios that serve as diagnostics for elemental C/H (e.g., \ch{CO_2}/\ch{H_2O}), O/H (e.g., \ch{H_2O}/\ch{C_2H_2}) abundances, thus indirectly the C/O ratio. These tracers are based on species and spectral regions accessible to JWST/MIRI observations.
    \item Application to observations: When compared with recent observations of T Tauri disks with confirmed dust gaps, our diagnostic flux ratios reveal evidence of varying carbon- and oxygen-enrichment across sources (below C/O of 1), highlighting the role of transport processes such as radial drift and disk gaps.
\end{enumerate}
This study emphasizes the need to consider both absolute elemental abundances and their ratios when interpreting disk chemistry. While our models isolate the effects of elemental abundances, future work should extend this approach by simultaneously varying key disk properties, such as inner disk radius and gaps, gas-to-dust ratio, dust settling, and UV radiation field, which evolve over time and significantly influence the mid-infrared emission (see \citealt{2026Esteve}).

\begin{acknowledgements}
We thank the Center for Information Technology of the University of Groningen for their support and for providing access to the H\'abr\'ok high performance computing cluster. I.K., A.M.A., and E.v.D. acknowledge support from grant TOP-1 614.001.751 from the Dutch Research Council (NWO). T.K. acknowledges support from STFC Grant ST/Y002415/1.
\end{acknowledgements}

  \bibliographystyle{aa} 
  \bibliography{ref.bib} 

\appendix
\onecolumn
\section{Fiducial model and non-LTE OH data}
\label{app:OH}
The disk model parameters that remain common through out the grid are listed in Table\,\ref{tab:propertiesTable}. \hl{The physical structure of the disk follows the model of \citet{woitke2016consistent}, incorporating the smooth inner disk edge and density structure from Eq. 71 of \citet{2024A&A...683A.219W}. Including this smooth edge is crucial for accurately predicting mid-infrared emission from the inner disk \citep{2024A&A...683A.219W,2024PASP..136e4302H}.}

In this work, we use an OH line list, encompassing levels and radiative transitions for $\nu$=0 and $\nu$=1 ($N\leq54$ and $N\leq50$ for $\nu$=0 and 1, respectively), based on \citet{2016JQSRT.168..142B}. The collision rates with o-\ch{H_2} and p-\ch{H_2} between pure rotational levels in the LAMDA database \citep{2005A&A...432..369S} served as the basis to estimate all the missing rates. The methodology follows \citet{2024A&A...691A..11T} for the expansion of the $\nu$=0 rates. Since we considered collisional rates between normal \ch{H_2} and OH, the rates with o-\ch{H_2} and p-\ch{H_2} have been merged into a single set of normal-\ch{H_2} assuming a constant \ch{H_2} ortho-to-para ratio of 3:1.  The rates have been further extrapolated for transitions arising from higher rotational levels and temperatures up to 2000\,K assuming that rates $\gamma \propto {\rm exp}(-\Delta E/kT)$, where $\Delta E$ is the energy difference between two levels, $k$. The collision rates from levels arising from $\nu$=1 were computed using the equal probability method and the coefficients in \citet{1999ApPhB..69...61R}. In this study, we did not consider other potential collision partners.

\begin{table}[!htbp]
\centering
\caption{Common disk model parameters. }

\begin{minipage}{0.45\linewidth}
\centering
\begin{tabular}{p{0.5\linewidth} p{0.2\linewidth} p{0.25\linewidth}}
\toprule
\textbf{Property} & \textbf{Symbol} & \textbf{Value} \\
\midrule
Stellar Mass          & $M_{*}$ & 0.7 $M_{\odot}$ \\
Effective temperature & $T_{*}$ & 4000 K  \\
Stellar luminosity    & $L_{*}$ & 1 $L_{\odot}$ \\ 
UV excess             & $f_{\rm UV}$ & 0.01           \\
UV powerlaw index     & $p_{\rm UV}$ & 1.3     \\
X-ray luminosity      & $L_{\rm x}$ & 10$^{30}$ erg/s \\
X-ray emission temperature & $T_{\rm x}$ & 2$\times$10$^7$ K \\ \hline
Strength of interstellar UV  & $\chi^{\rm ISM}$ & 1 \\
Cosmic ray H$_2$ ionization rate & $\zeta_{\rm CR}$ & 1.7$\times$10$^{-17}$ s$^{-1}$  \\ \hline
Disk mass  & $M_{\rm disk}$ & 0.02 $M_{\odot}$ \\
Gas-to-dust ratio  & $M\rm _g$/M$\rm _d$ & 100 \\
Inner disk radius  & $R_{\rm in}$ & 0.05 au \\
Tapering-off radius & $R_{\rm tap}$ & 30 au \\
Column density power index  & $\hl{\eta}$ & 1 \\
\bottomrule
\end{tabular}
\end{minipage}
\hfill
\begin{minipage}{0.45\linewidth}
\centering
\begin{tabular}{p{0.5\linewidth} p{0.2\linewidth} p{0.2\linewidth}}
\toprule
\textbf{Property} & \textbf{Symbol} & \textbf{Value} \\
\midrule
Reference scale height  & $H_{\rm g}$(100 au) & 10 au \\  
Flaring power index     & $\beta$ & 1.15 \\  
Smooth edge parameter 1 & reduc & 10$^{-6}$ \\
Smooth edge parameter 2 & raduc & 1.5 \\ \hline
Minimum dust \hl{grain} radius   & $a_{\rm min}$ & 0.05 $\mu$m \\ 
Maximum dust \hl{grain} radius   & $a_{\rm max}$ & 3000 $\mu$m \\ 
Dust size dist. power index & $a_{\rm pow}$ & 3.5 \\ 
Turbulent mixing parameter & $\alpha_{\rm settle}$ & 0.001 \\ 
Refractory dust composition & Mg$_{0.7}$Fe$_{0.3}$SiO$_3$ & 60\% \\
                          & amorph. C & 15\% \\
                          & porosity & 25\% \\ 
\hline
Chemical heating efficiency & $\gamma^{\rm chem}$ & 0.2 \\ 
Distance to the observer & $d$ & 140 pc \\ 
\bottomrule
\end{tabular}
\end{minipage}
\tablefoot{Parameter definitions can be found in \citet{woitke2016consistent} and \citet{2024A&A...683A.219W}.}
\label{tab:propertiesTable}
\end{table}

\section{Heating and cooling}
\begin{figure}[!htbp]
    \centering
    \includegraphics[width=0.99\linewidth]{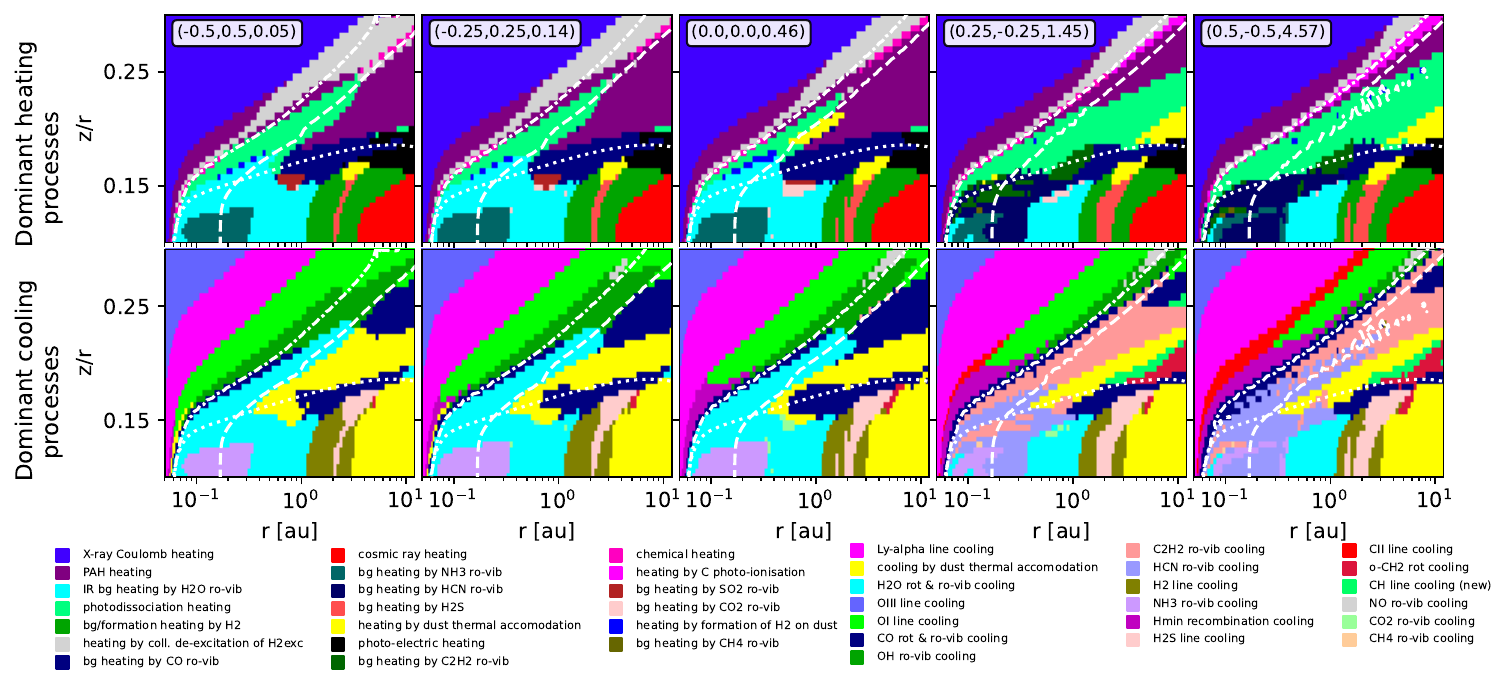}
    \caption{Dominant heating and cooling processes across models of different C/O ratios. The white dashed lines indicate the 300\,K and 1000\,K gas temperature contours. The white dotted lines indicate the A$\rm _v$=1 line.}
    \label{fig:multi_heatcool}
\end{figure}

A comparison of the dominant heating and cooling processes for five models of different C/O ratios are presented in Fig.\,\ref{fig:multi_heatcool}.

\section{Diverse spectral appearance with the same C/O ratio}
\label{sec:sameco}

\begin{figure}[!ht]
    \includegraphics[width=0.99\linewidth]{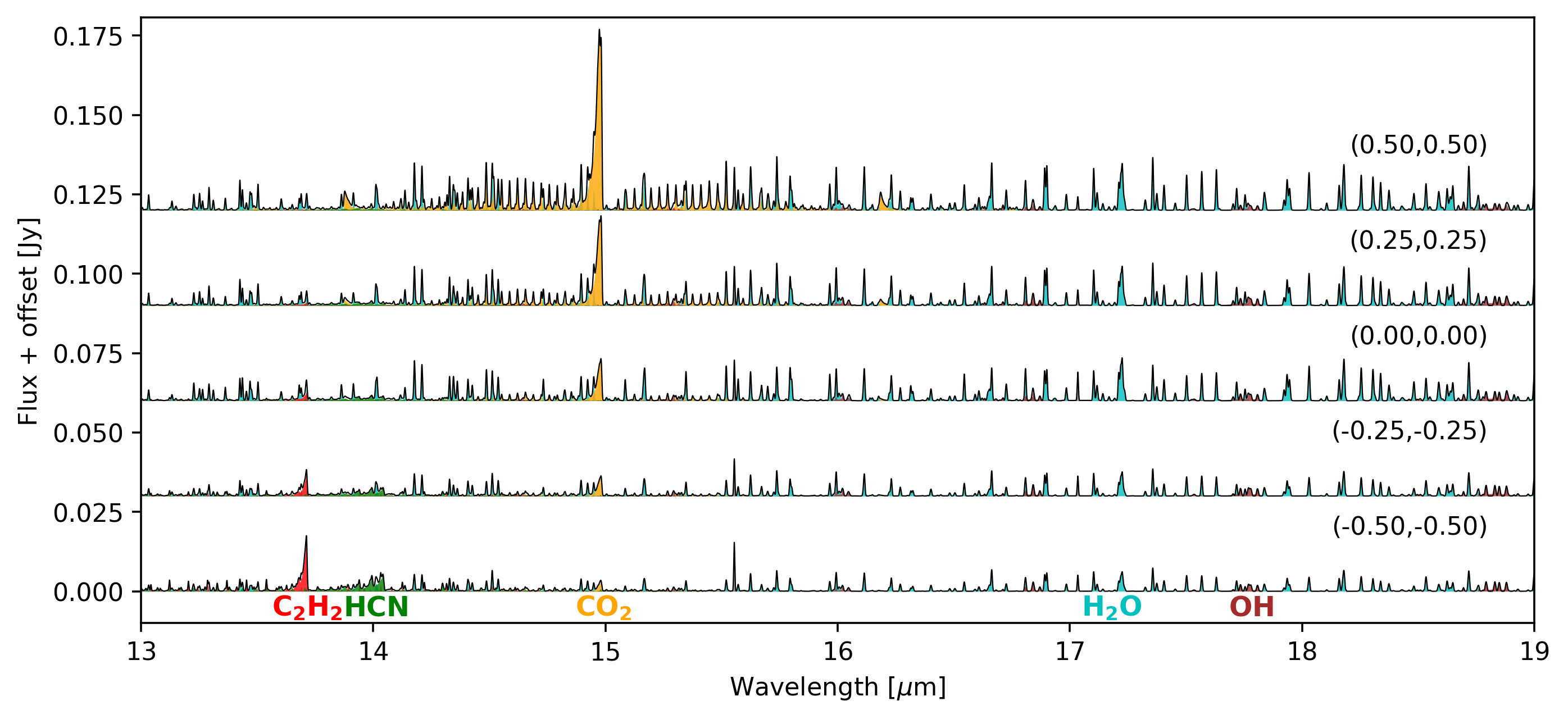}
    \caption{FLiTs spectra of five models with the same C/O ratio of 0.45, but different C/H and O/H (indicated in the figure). The spectrum in the center is the fiducial model (0,0).}
    \label{fig:co_change}
\end{figure}

\begin{figure*}[!ht]
    \centering
    \includegraphics[width=0.99\linewidth]{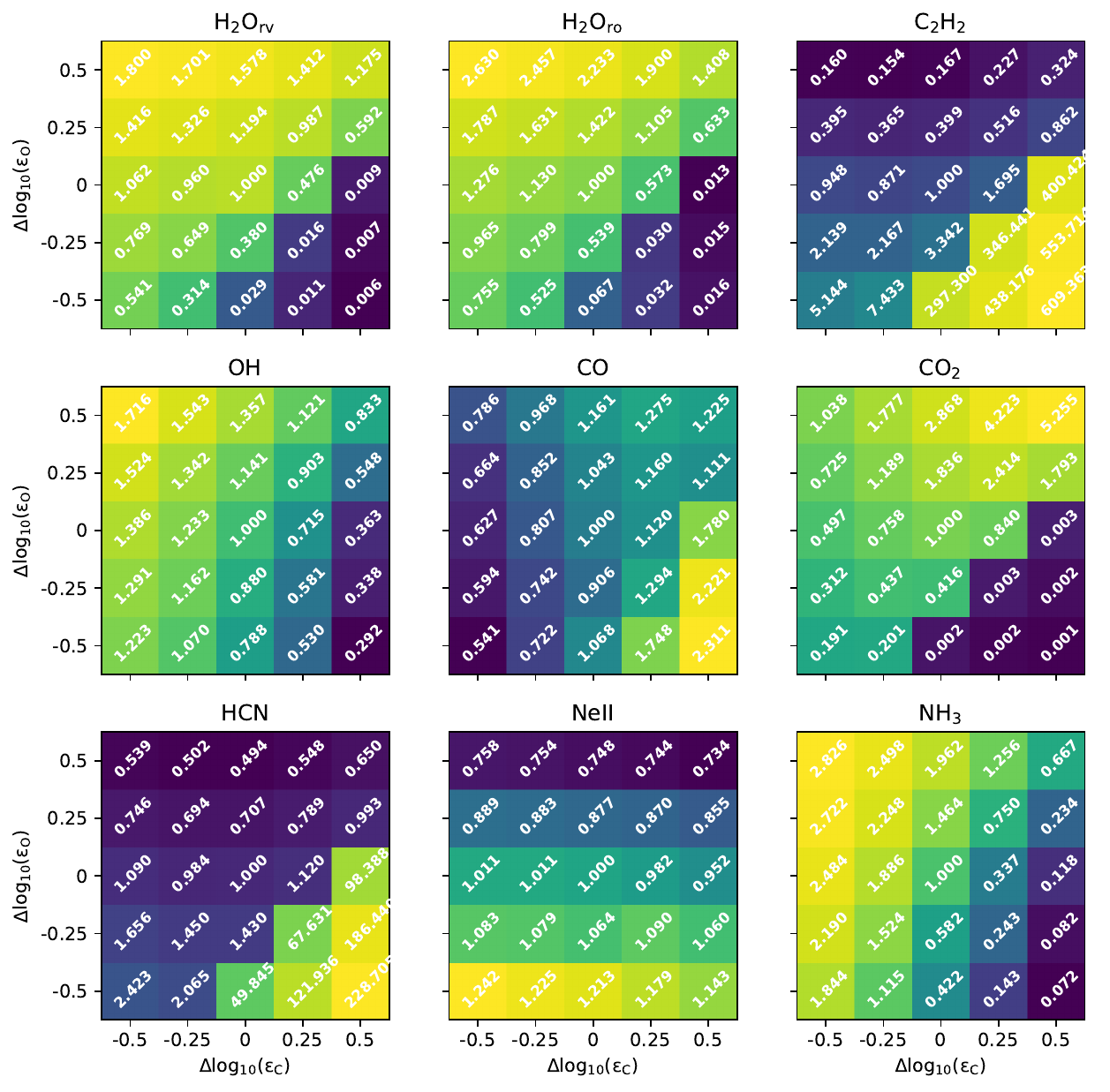}
    \caption{Integrated fluxes normalized to the fiducial model.}
    \label{fig:gdr_100_fluxsums}
\end{figure*}

Figure\,\ref{fig:co_change} shows spectra of five models from the grid that have the same C/O ratio but different abundances of carbon and oxygen. The spectra are diverse, varying between spectra dominated by \ch{CO_2}, \ch{H_2O}, or \ch{C_2H_2}, all for the same C/O ratio. Figure\,\ref{fig:gdr_100_fluxsums} presents a quantitative comparison of integrated molecular fluxes of different species, each normalized to the flux from the fiducial model.

\section{Abundances and emitting regions different species}
\label{sec:miscmolecules}
\begin{figure}[!ht]
    \centering
    \includegraphics[width=0.99\linewidth]{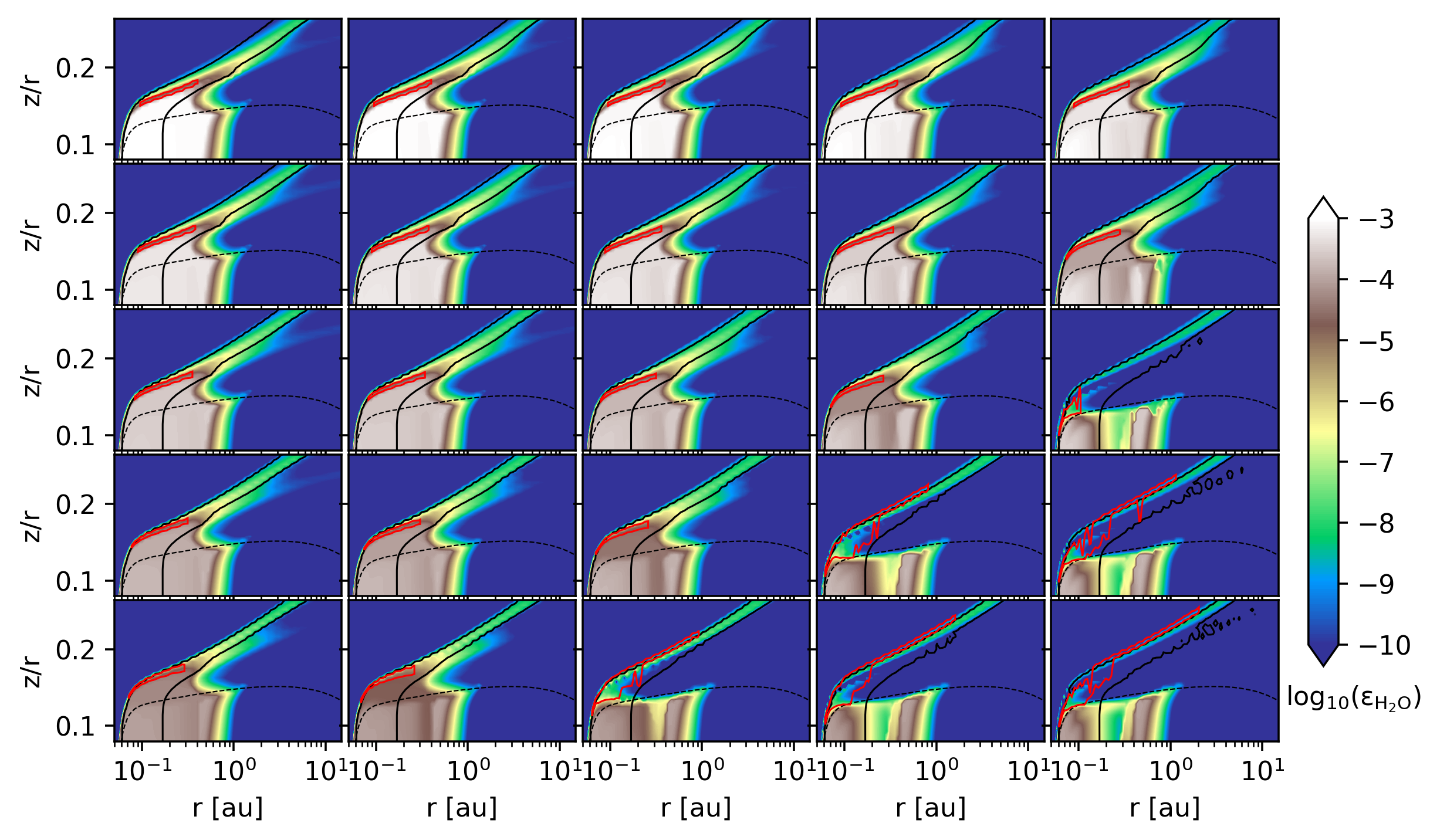}
    \caption{Abundance and line emitting region (red line) of \ch{H_2O}. The solid black lines indicate the gas temperatures of 300\,K and 1000\,K, the dotted line indicates the $\rm \tau_{H_2O}$=1 line. Each panel corresponds to the same model illustrated in Fig.\ref{fig:co_grid}.}
    \label{fig:lineemittingregions_h2o}
\end{figure}

\begin{figure}[!ht]
    \centering
    \includegraphics[width=0.99\linewidth]{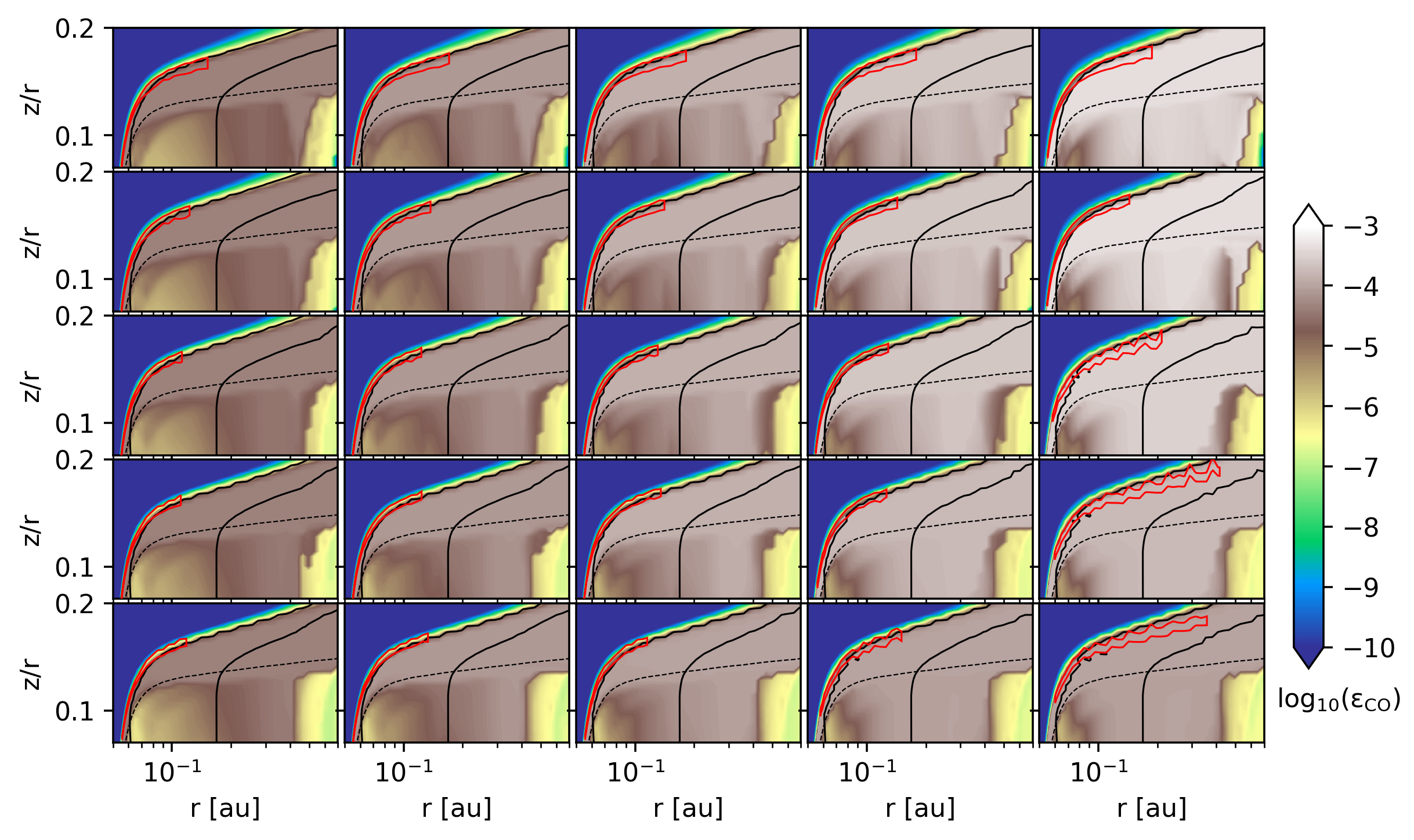}
    \caption{Same as Fig.\ref{fig:lineemittingregions_h2o}, but for \ch{CO}. The dashed line indicates the $\rm \tau_{CO}$=1 line.}
    \label{fig:lineemittingregions_co}
\end{figure}

\begin{figure}[!ht]
    \centering
    \includegraphics[width=0.99\linewidth]{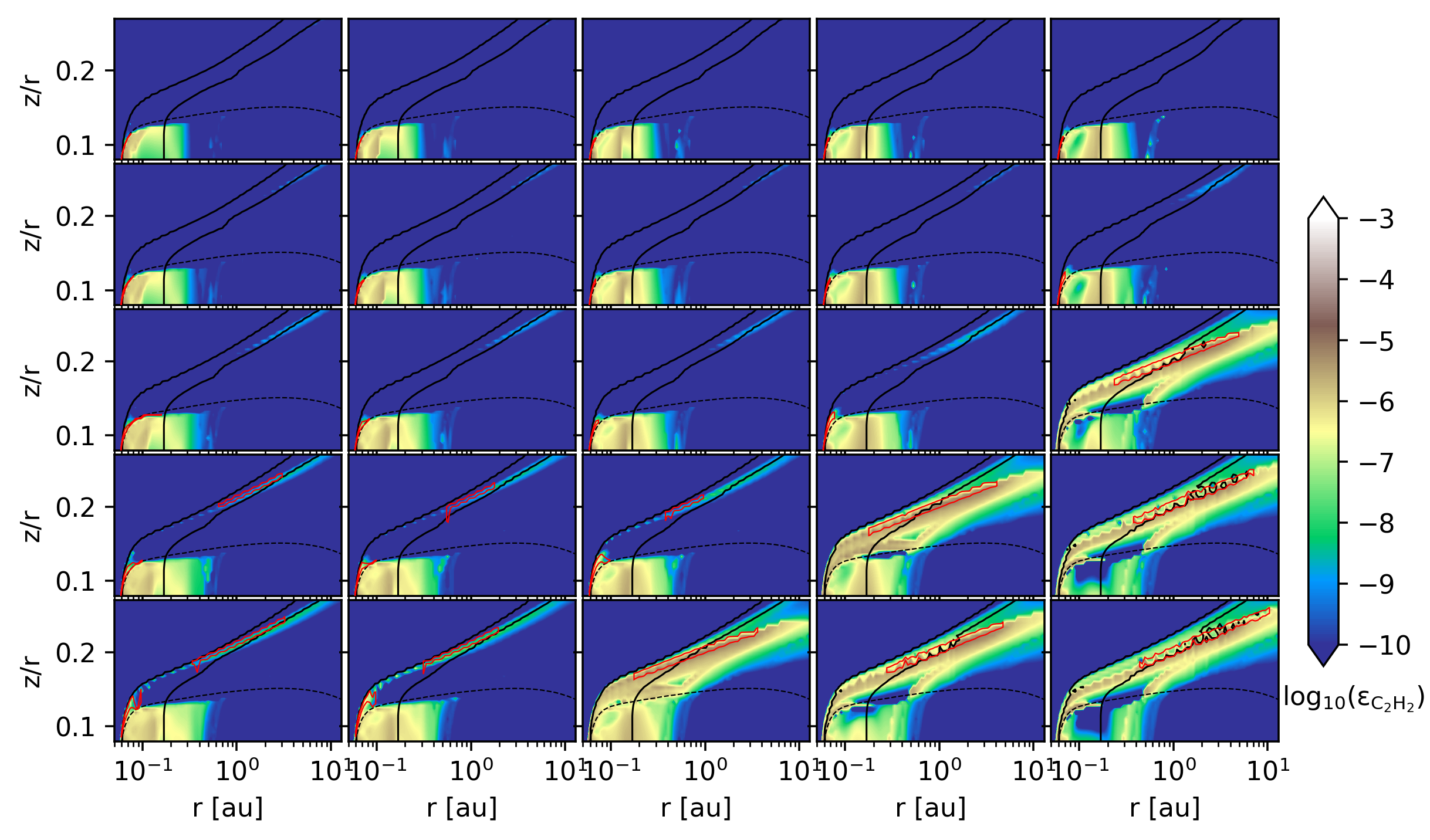}
    \caption{Same as Fig.\ref{fig:lineemittingregions_h2o}, but for \ch{C_2H_2}. The dashed line indicates the $\rm \tau_{C_2H_2}$=1 line.}
    \label{fig:lineemittingregions_c2h2}
\end{figure}

\begin{figure}[!ht]
    \centering
    \includegraphics[width=0.99\linewidth]{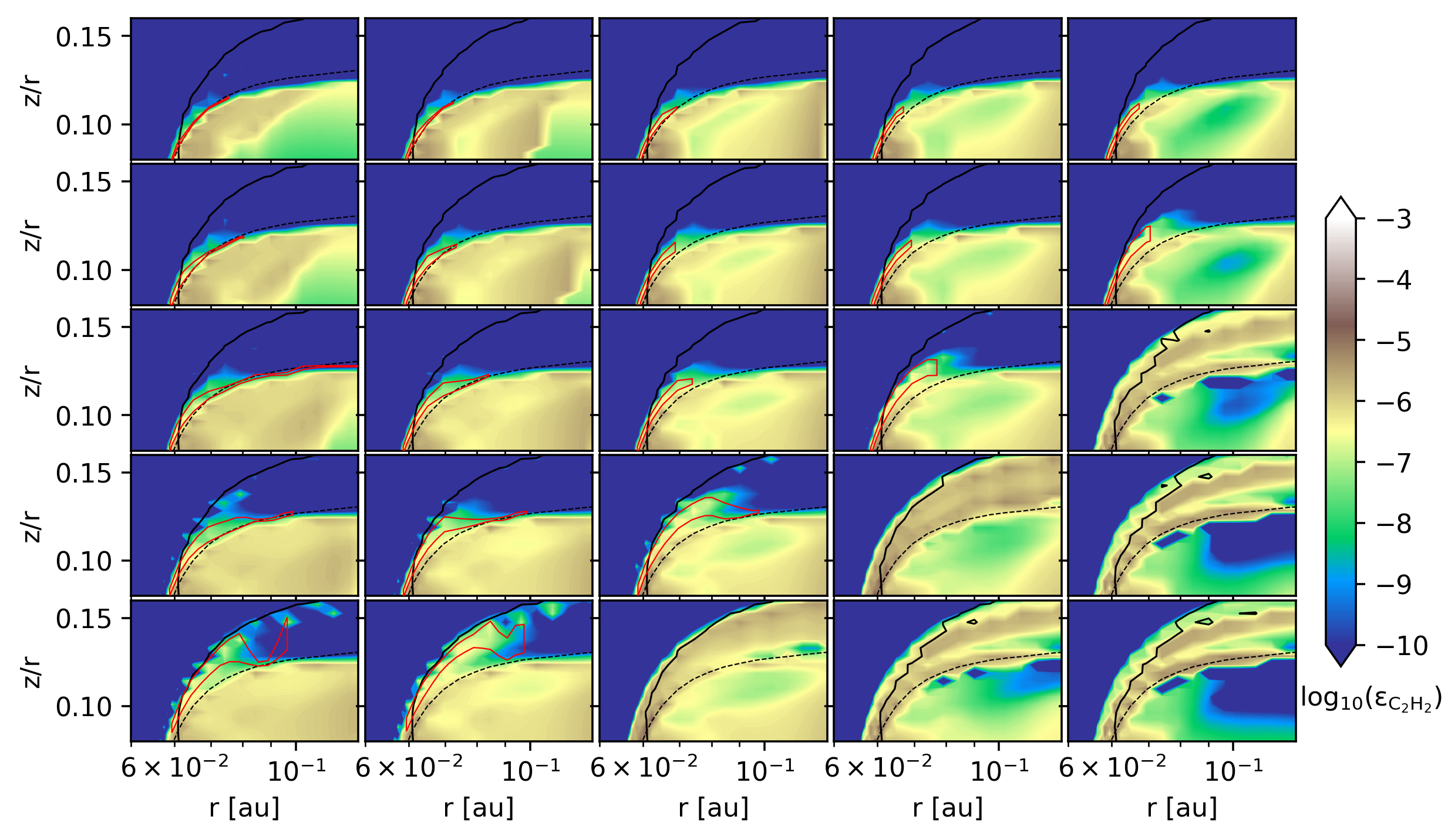}
    \caption{Same as Fig.\ref{fig:lineemittingregions_c2h2}, but zoomed-in.}
    \label{fig:lineemittingregions_c2h2_zoom}
\end{figure}

\begin{figure}[!ht]
    \centering
    \includegraphics[width=0.99\linewidth]{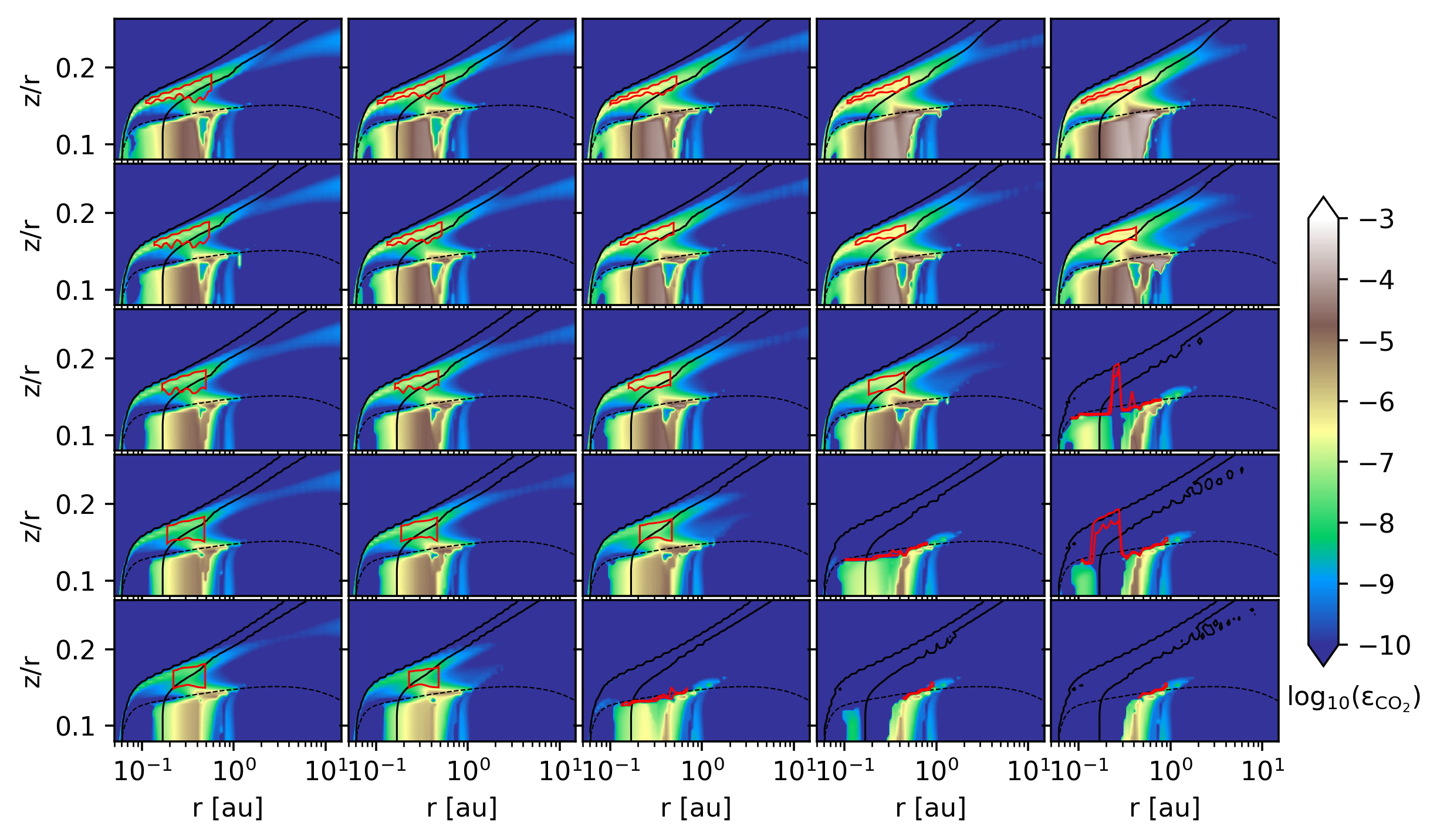}
    \caption{Same as Fig.\ref{fig:lineemittingregions_h2o}, but for \ch{CO_2}. The dashed line indicates the $\rm \tau_{CO_2}$=1 line.}
    \label{fig:lineemittingregions_co2}
\end{figure}

\begin{figure}[!ht]
    \centering
    \includegraphics[width=0.99\linewidth]{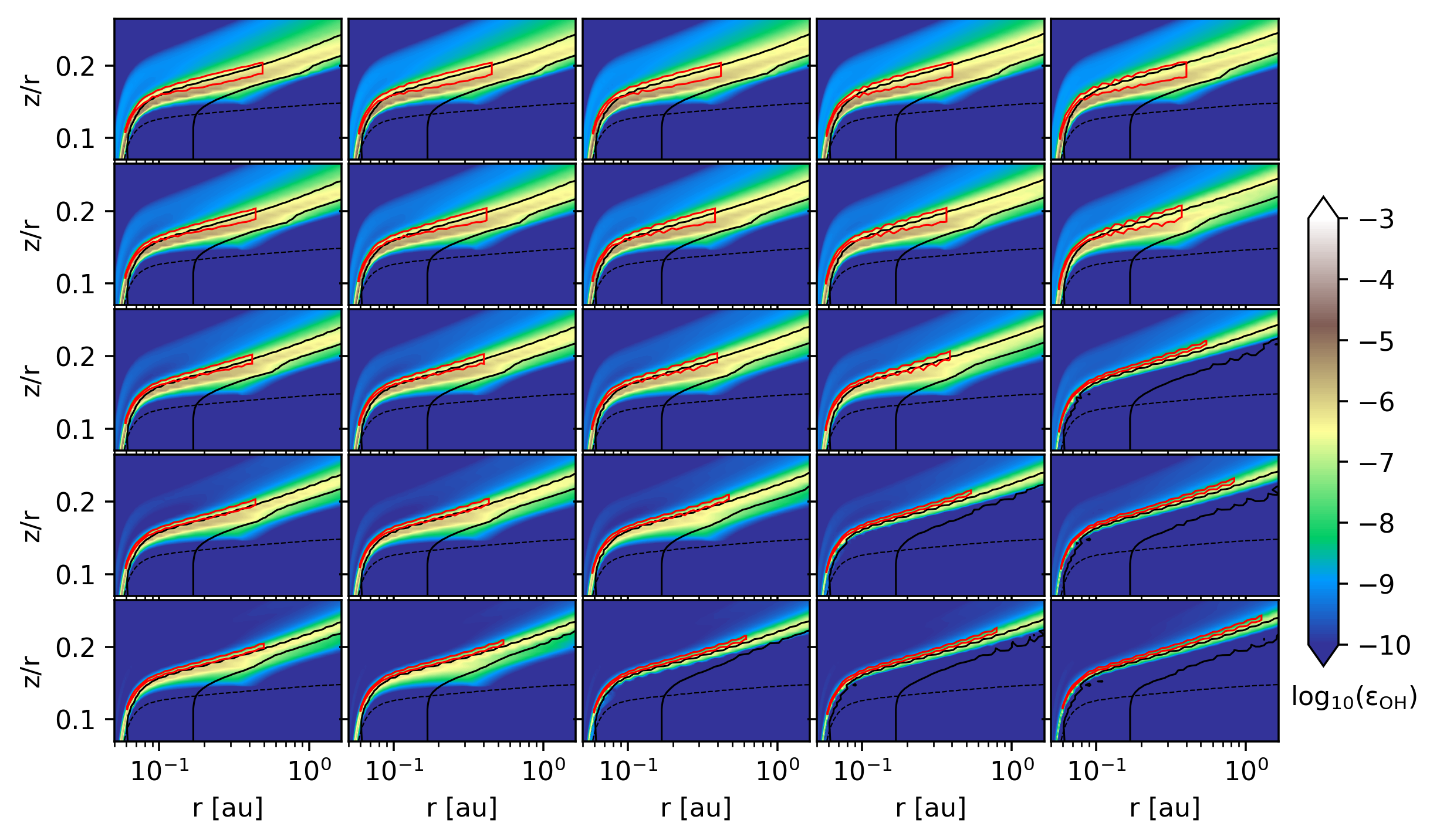}
    \caption{Same as Fig.\,\ref{fig:lineemittingregions_h2o}, but for \ch{OH}. The dashed line indicates the $\rm \tau_{OH}$=1 line.}
    \label{fig:lineemittingregions_oh}
\end{figure}

\begin{figure}[!ht]
    \centering
    \includegraphics[width=0.99\linewidth]{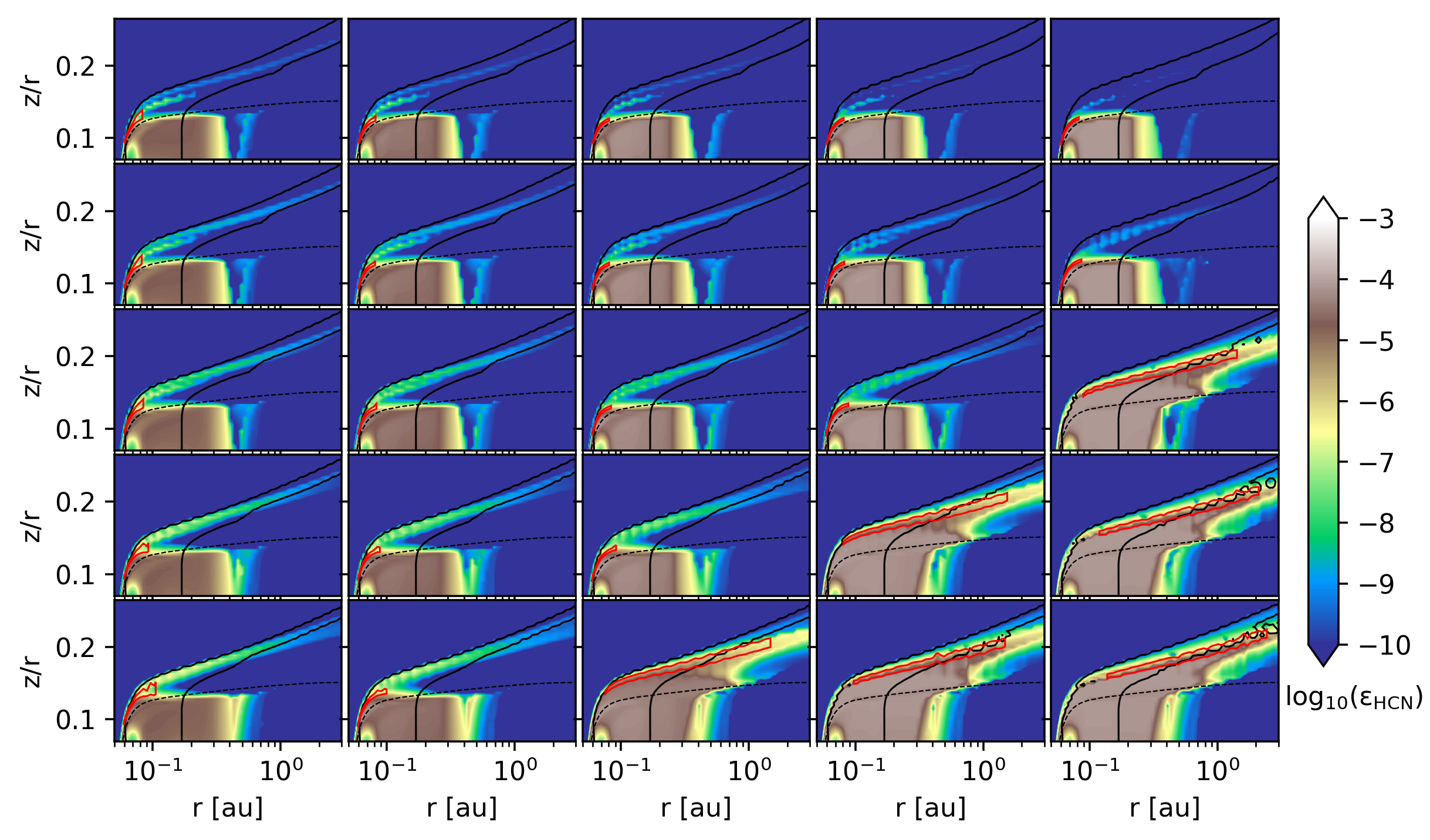}
    \caption{Same as Fig.\,\ref{fig:lineemittingregions_h2o}, but for \ch{HCN}. The dashed line indicates the $\rm \tau_{HCN}$=1 line.}
    \label{fig:lineemittingregions_hcn}
\end{figure}

\begin{figure}[!ht]
    \centering
    \includegraphics[width=0.99\linewidth]{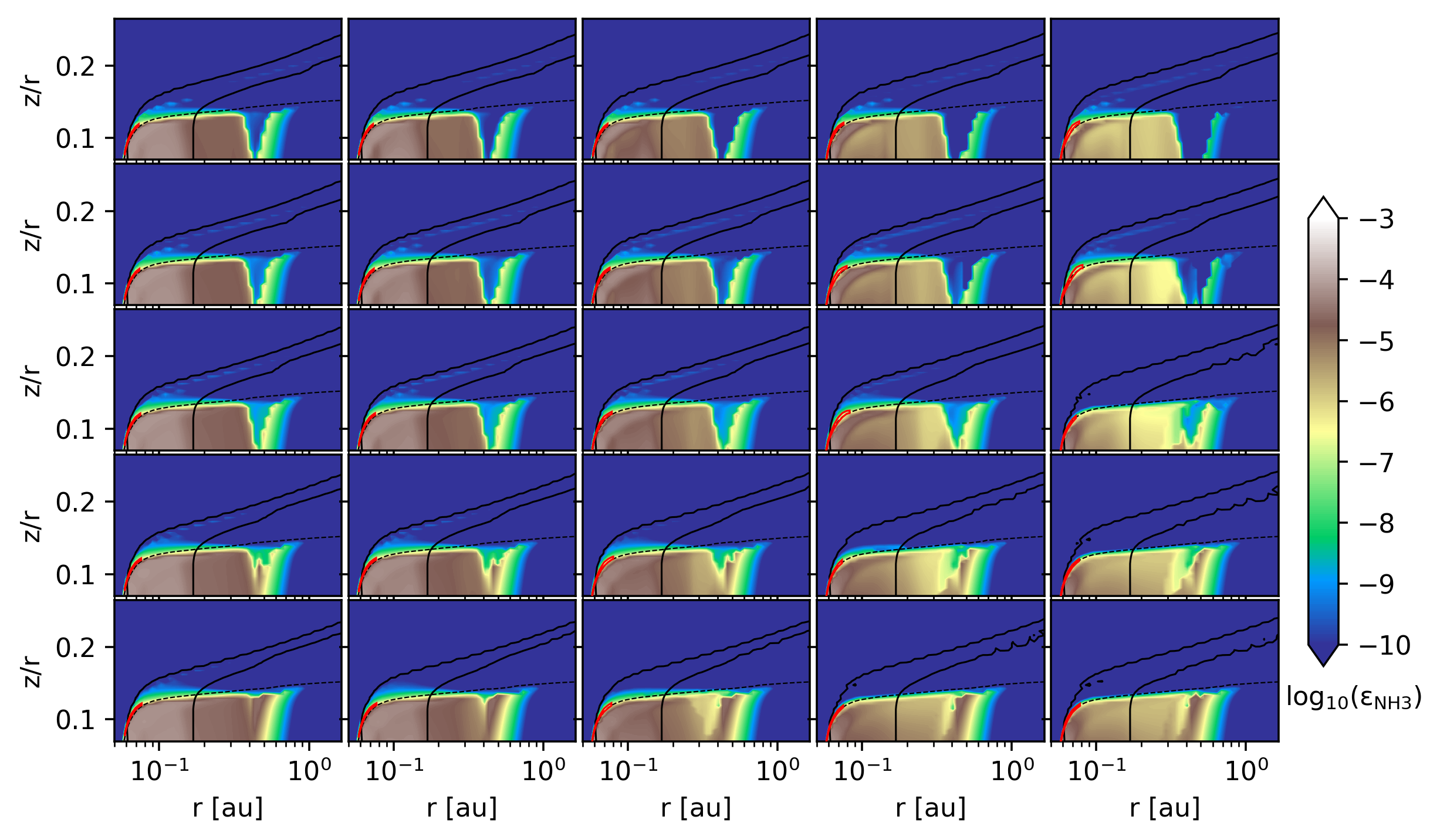}
    \caption{Same as Fig.\,\ref{fig:lineemittingregions_h2o}, but for \ch{NH_3}. The dashed line indicates the $\rm \tau_{NH_3}$=1 line.}
    \label{fig:lineemittingregions_nh3}
\end{figure}

\begin{figure}[!ht]
    \centering
    \includegraphics[width=0.99\linewidth]{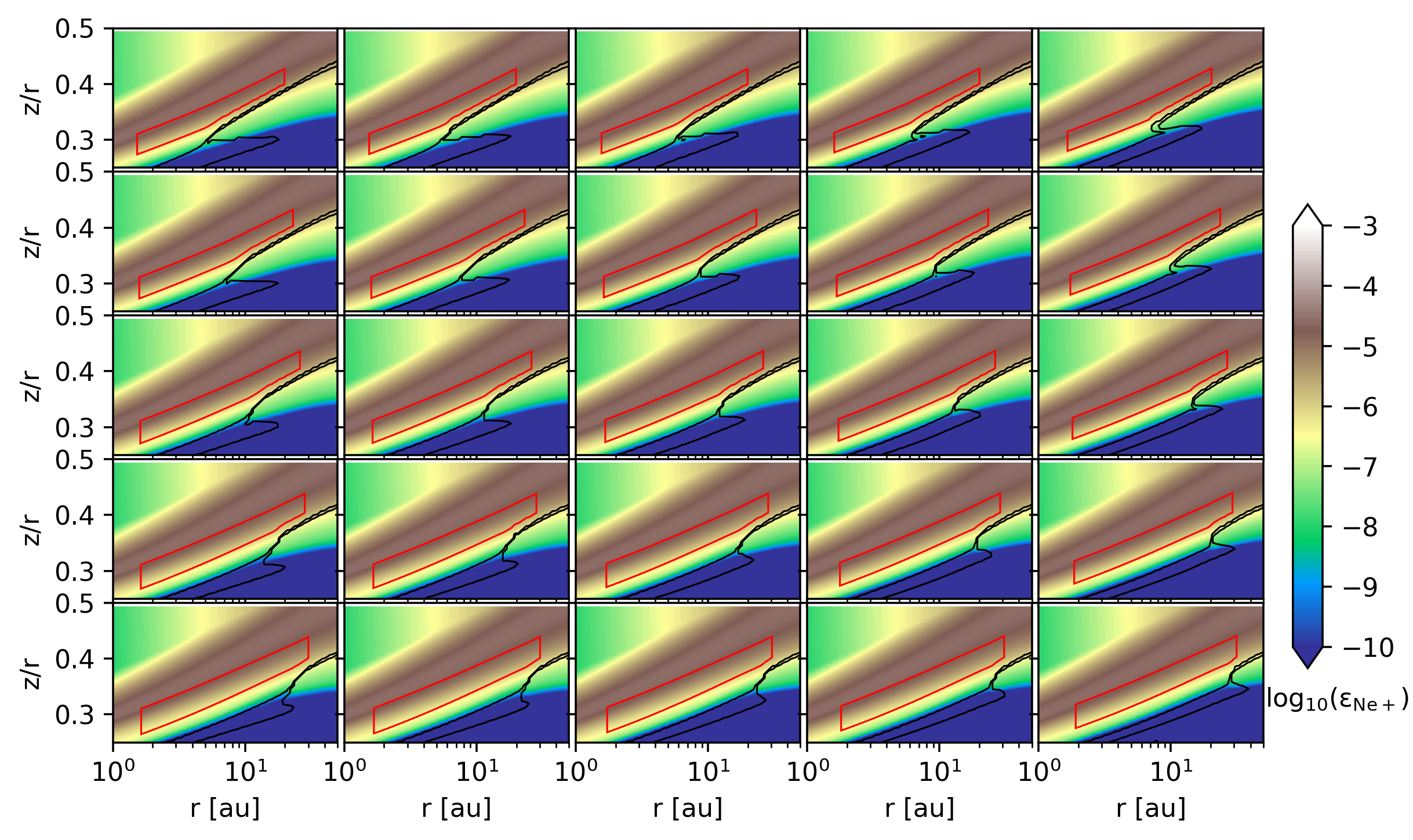}
    \caption{Same as Fig.\,\ref{fig:lineemittingregions_h2o}, but for \ch{Ne^{+}}. The dashed line indicates the $\rm \tau_{[Ne II]}$=1 line. Note the different axes limits.}
    \label{fig:lineemittingregions_ne+}
\end{figure}

Figures\,\ref{fig:lineemittingregions_h2o}-\ref{fig:lineemittingregions_ne+} show the abundances of the species discussed in this paper. \ch{H_2O}, \ch{CO}, \ch{C_2H_2}, and \ch{CO_2} are discussed in the main text (Sect.\,\ref{sec:generalappearance}). \hl{The line-emitting regions are bounded by four contours. The upper and lower boundaries indicate the 15\% and 85\% levels of the local vertical flux (integrated from the surface to the midplane). The vertical lines mark the radii where the radially cumulative flux reaches 15\% and 85\% of the total emission. Together, these boundaries enclose approximately 50\% of the total line flux \citep[also see][]{woitke2016consistent}. We calculate the flux contribution of individual grid points using the escape probability formalism presented in Sect. 2.2 of \citet{2024A&A...683A.219W}.}

Figure\,\ref{fig:lineemittingregions_oh} shows the abundances and the line emitting regions of \ch{OH}. Clearly, higher oxygen abundances lead to higher \ch{OH} abundances. The emitting region morphs from a thin-long layer to a thick-narrow layer with increasing oxygen abundances. The temperature of the emitting region also moves slightly higher up and lower down around the 1000\,K contour with changing oxygen abundance.

Figure\,\ref{fig:lineemittingregions_hcn} shows \hl{that} the abundances and the line emitting regions of \ch{HCN} behave similar to \ch{C_2H_2}. When C/O$<$1, the emitting region is a narrow layer above the $\tau$=1 line. The abundances in the surface layer increase with decreasing oxygen abundance. The abundances increases significantly when the C/O ratio is greater than 1, and the emitting region extends to much larger areas.

Figure\,\ref{fig:lineemittingregions_nh3} shows the abundances and the line-emitting regions of \ch{NH_3}. The \ch{NH_3} reservoir is largely limited to the region below the $\tau$=1 layer. The low abundances above this layer and the very narrow emitting region lead to small fluxes. 

Figure\,\ref{fig:lineemittingregions_ne+} shows the abundances and the line-emitting regions of \ch{[Ne II]}. The \ch{Ne^{+}} reservoir is largely limited to the high temperature region above 1000\,K. Since \ch{Ne^{+}} is inert, the abundances do not change with carbon and oxygen abundances. However, the emitting regions change slightly in response to changes in the temperature structure of the gas.

\section{C/O diagnostics with very low-mass stars}
\label{sec:vlms_obs}
\begin{figure*}[!ht]
    \centering
    \sidecaption
    \includegraphics[width=12cm]{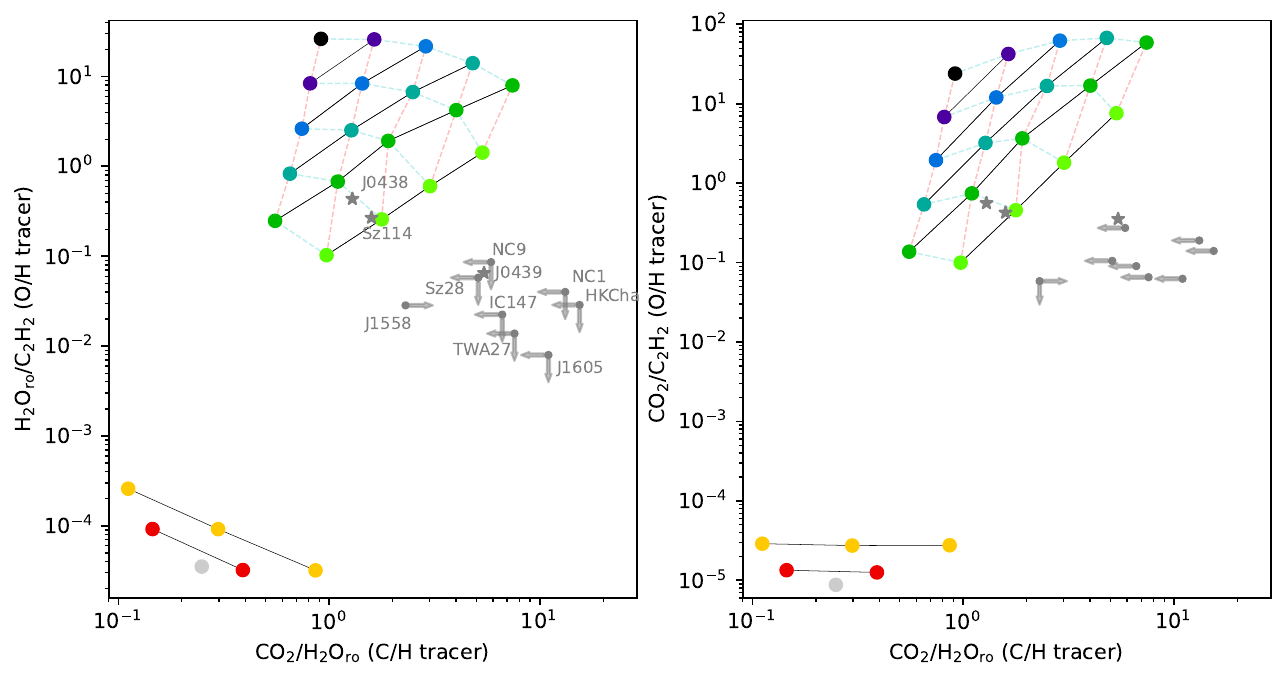}
    \caption{Same as Fig.\,\ref{fig:obs_c_o_tracer} but for observations of disks around very low-mass stars, and the wavelength intervals are adapted for these sources. The spectra and the detections are taken from \citet{2025A&A...699A.194A}, \citet{2025ApJ...984L..62A} and \citet{2023ApJ...959L..25X}.}
    \label{fig:obs_vlms_c_o_tracer_2}
\end{figure*}

Similar to Fig.\,\ref{fig:obs_c_o_tracer}, Fig.\,\ref{fig:obs_vlms_c_o_tracer_2} presents the observed molecular flux ratios of disks around 11 very low-mass stars, which are generally carbon-rich, in comparison with those of the models. These sources are taken from \citet{2025A&A...699A.194A} with one source from \citet{2023ApJ...959L..25X}. The observations generally fall outside the flux ratios traced by the models. J0438 (highly inclined) and Sz114 have a water-rich MIRI spectra \citep{2026ApJ...997..281P,2025ApJ...984L..62A,2023ApJ...959L..25X}. Comparing the model flux ratios to the observed flux ratios of J0438 and Sz114 indicates C/O ratios less than 1. All the other sources have multiple bright hydrocarbon emission corresponding to C/O ratios greater than 1 \citep{2025A&A...699A.194A,2024A&A...689A.231K}. In fact, \citet{2026A&A...705A.222K} show that models require a C/O ratio significantly larger than unity to reproduce the \ch{C_2H_2} pseudo-continuum observed in J1605. Moreover, not much is known about the inner and outer disk structures of these carbon-rich sources.

\begin{figure*}[!ht]
    \centering
    \sidecaption
    \includegraphics[width=12cm]{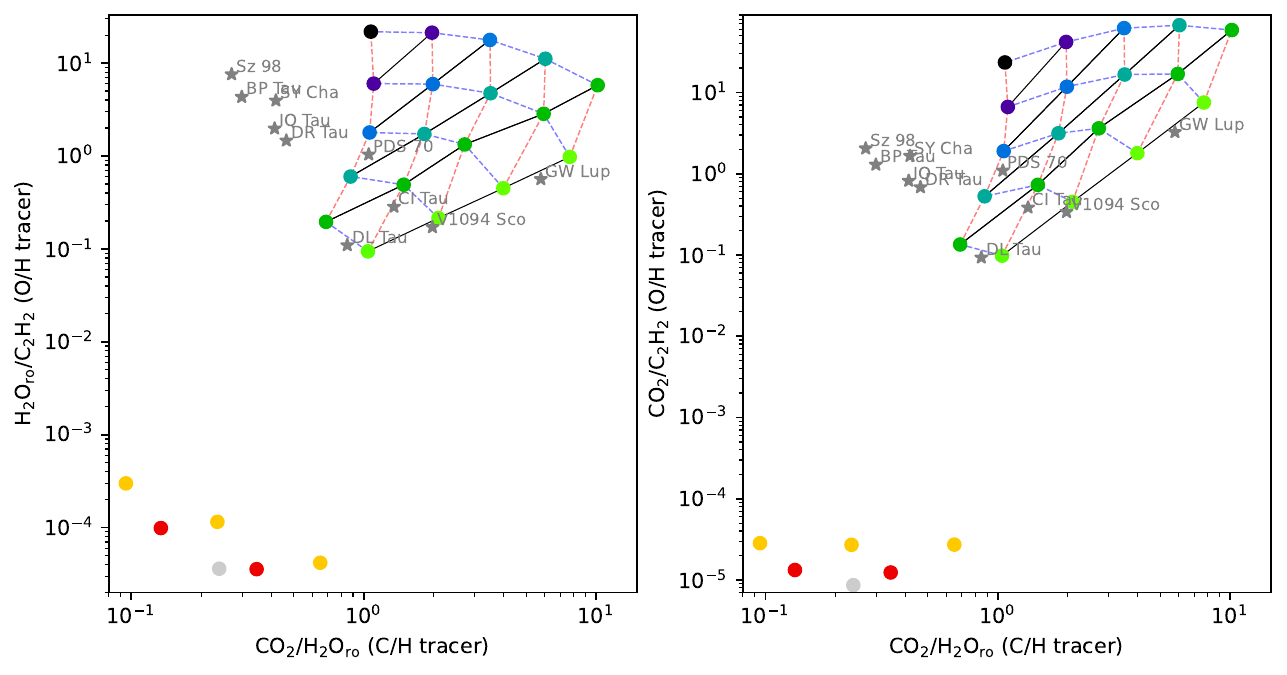}
    \caption{Same as Fig.\,\ref{fig:obs_c_o_tracer} but includes models with C/O$>$1. The colors of the points correspond to the C/O ratios (see Fig.\,\ref{fig:gdr_100_fluxratios_CO_diagnostics}). The models with C/O$>$1 are at the bottom left corner of each panel.}
    \label{fig:obs_c_o_tracer_co_1}
\end{figure*}

\end{document}